\documentclass[sigconf]{acmart}

\AtBeginDocument{%
  }
\usepackage{subfigure}
\usepackage{adjustbox}
\usepackage{algorithm}
\usepackage[noend]{algorithmic}

\newcommand\sxrightarrow[1]{%
    \raisebox{0.2ex}{\adjustbox{trim=0pt 2pt 0pt 0pt}{$\xrightarrow{\raisebox{-.5ex}[0pt][0pt]{\ensuremath{\scriptstyle#1}}}$}}%
}

\newcommand{\xtwoheadrightarrow}[2][]{%
    \xrightarrow[#1]{#2}\mathrel{\mkern-14mu}\rightarrow
}

\newcommand\sxtwoheadrightarrow[1]{%
    \raisebox{0.2ex}{\adjustbox{trim=0pt 2pt 0pt 0pt}{$\xtwoheadrightarrow{\raisebox{-.5ex}[0pt][0pt]{\ensuremath{\scriptstyle#1}}}$}}%
}

\begin{document}

\title{Vbox: Efficient Black-Box Serializability Verification}

\author{Weihua Sun}
\authornotemark[1]
\affiliation{%
  \institution{Harbin Institute of Technology}
  \city{Harbin}
  \state{Heilongjiang}
  \country{China}
}

\author{Zhaonian Zou}
\authornotemark[2]
\affiliation{%
  \institution{Harbin Institute of Technology}
  \city{Harbin}
  \state{Heilongjiang}
  \country{China}
}



\begin{abstract}
Verifying the serializability of transaction histories is essential for users to know if the DBMS ensures the claimed serializable isolation level without potential bugs. Black-box serializability verification is a promising approach. Existing verification methods often have one or more limitations such as incomplete detection of data anomalies, long verification time, high memory usage, or dependence on specific concurrency control protocols. In this paper, a new black-box serializability verification method called \textsf{Vbox} is proposed. \textsf{Vbox} is powered by a number of new techniques, including the support for predicate database operations, comprehensive applications of transactions' time information in the verification process, and a simplified satisfiability (SAT) problem formulation and its efficient solver. In this paper, \textsf{Vbox} is verified to be correct, efficient, and capable of detecting more data anomalies, while not relying on any specific concurrency control protocols.
\end{abstract}




\maketitle

\section{Introduction}

Transaction processing is fundamental to database management systems (DBMS), ensuring data consistency and persistency~\cite{GrayR93}. A key aspect is isolation~\cite{cerone_framework_2015,Crooks_Pu_Alvisi_Clement_2017}, which prevents interference among concurrent transactions. The highest isolation level, ``serializable,'' guarantees that the concurrent execution of transactions is equivalent to a sequential execution~\cite{BernsteinHG87,Papadimitriou79b}. Despite DBMSs claiming serializable isolation, anomalies occur~\cite{cui_understanding_2024,dou_detecting_2023,kingsbury_elle_2020}, indicating potential implementation bugs. Thus, it is necessary for users or programmers to verify that the transaction histories are serializable. Black-box verification methods regard the DBMS as a black-box and directly verify serializability by checking the histories itself.

However, black-box serializability verification is NP-complete~\cite{BernsteinG83}. \textsf{Cobra}~\cite{tan_cobra_2020} constructs a direct serializable graph (DSG)~\cite{adya_generalized_2000} from client-collected information and checks for cycles. If the DSG is acyclic, the transaction history is serializable. 

Another approach verifies if the transaction history conforms to specific concurrency control protocols. \textsf{Leopard}~\cite{li_leopard_2023} records operation timestamps from clients to infer execution order. If the execution order violates the concurrency control protocol (e.g., serializable snapshot isolation (SSI)~\cite{ports_serializable_2012,FeketeLOOS05} disallows concurrent updates to the same object), \textsf{Leopard} regards the history non-serializable.

A third approach check the existence a valid commit order for the transactions. Biswas and Enea~\cite{biswas_complexity_2019} propose \textsf{BE}, assumes that transaction history is organized by sessions, and the transactions in each session follow the session order. The \textsf{BE} algorithm selects transactions from the head of each session and adds them to the commit order, ensuring that each selected transaction does not depend on the transactions already in the commit order.

Existing methods have intrinsic limitations. \textsf{Cobra} and \textsf{BE} lack support for predicate operations. In real-world applications, predicate operations are common and may cause special data anomalies.  Additionally, the scalability of both \textsf{Cobra} and \textsf{BE} is limited. Particularly, the time complexity of \textsf{BE} is $O(n^{s + 3})$, where $n$ is the number of transactions, and $s$ is the number of sessions. The verification time and the memeory overhead of \textsf{Cobra} grow sharply, even exponentially, with the number of transactions.

\textsf{Leopard}'s verification is neither universal nor complete. It cannot infer the execution order for protocols such as optimistic concurrency control (OCC)~\cite{KungR81}, timestamp ordering (TO)~\cite{BernsteinG81}, and Percolator~\cite{PengD10}, finally missing some anomalies that occur under these protocols. 

To address these limitations, we propose \textsf{Vbox}, a novel black-box serializability verifier based on Adya’s definition~\cite{adya_generalized_2000}. \textsf{Vbox} extends \textsf{Cobra}, inheriting its correctness and completeness while remaining protocol-agnostic. Our contributions include:

(1) We introduce predicate constraints to support DSGs with predicate dependency edges, enabling our method to handle transaction histories involving predicate read and write operations.

(2) We incorporate transactions' time information into the verification process. Specifically, the time information is applied to: (a) derive more dependency edges in the DSG, (b) add time dependency edges to the DSG as a supplement, (c) create a compact transitive closure structure and implement efficient graph algorithms to speed up DSG construction, and (d) provide heuristic guidance during DSG construction.

(3) We transform serializability verification into an satisfiability (SAT) problem and propose a simplified formulation of this problem.  We design a customized solver to speed up solving the simplified formulation of the SAT problem.

Our formal study and experimental evaluation verify that \textsf{Vbox} is
correct, efficient, and capable of detecting more data anomalies, while not
relying on any specific concurrency control protocols. (1) \textsf{Vbox} is
developed from \textsf{Cobra}. All the extensions to \textsf{Cobra} are proved
to be correct rigorously. (2) \textsf{Vbox} finds all the serializability
anomalies in the real-world and the synthetic transaction execution histories
containing predicate read and write operations, while the existing methods
cannot find all these anomalies. (3) \textsf{Vbox} has high verification
efficiency and low memory usage. For an execution history of 10K transactions,
\textsf{Vbox} is 60--100X faster than \textsf{Cobra} and has 20--70X lower
memory consumption. (4) \textsf{Vbox} exhibits nearly linear scalability. As
the number of transactions in the history increases from 10K to 100K, the
verification time of \textsf{Vbox} increases from 0.31s to 4.16s (13.4X), and
the memory usage increases from 44MB to 417MB (9.5X).

\section{Preliminaries}
\label{sec:preliminaries}

\noindent{\bf Databases.}
A database consists of a set of objects that can be read and written by transactions. Each object $x$ has one or more versions denoted as $x_1, x_2, x_3, \dots$. A read operation, denoted as $r(x, x_i)$, reads a specific version $x_i$ of object $x$, and a write operation, denoted as $w(x, x_j)$, creates a new version $x_j$ of object $x$. Let $\theta$ be a predicate, and $X$ the set of objects that can be evaluated by $\theta$. A predicate read operation based on $\theta$ can be represented as a set of $\theta$-reads $\{r_{\theta}(x, x_i) \mid x \in X\}$, where $r_{\theta}(x, x_i)$ reads version $x_i$ of object $x$ if $\theta(x_i)$ is true. Similarly, a predicate write operation is represented as a set of $\theta$-writes $\{w_{\theta}(x, x_i, x_j) \mid x \in X\}$, where $w_{\theta}(x, x_i, x_j)$ creates version $x_j$ of object $x$ if $\theta(x_i)$ is true.

\noindent{\bf Transactions.}
A transaction is a sequence of operations, starting with a \texttt{begin} operation, ending with a \texttt{commit} or \texttt{abort} operation, and in between are read operations (including predicate reads) and write operations (including predicate writes). During the course of transaction processing, new versions of objects are continuously created. The latest version of an object created by a committed transaction is called the \emph{installed version}. In this database model, a total order $<_v$ on all installed versions of an object is maintained, referred to as the \emph{version order}.

An \textit{complete history} of a set $T$ of transactions is defined as $(T,
  <_e, <_v)$, where $<_e$ is a partial order called the \emph{event order} on all
operations in $T$, and $<_v$ is the version order of the installed versions of
all objects in the database. From the user's perspective, $<_v$ is invisible,
and $x_i$ is unknown for any $\theta$-read $r_\theta(x, x_i)$ or $\theta$-write
$w_\theta(x, x_i, x_j)$ when $\theta(x_i) = \text{false}$. Such predicate
operations are denoted as $r_\theta(x, \_)$ or $w_\theta(x, \_, x_j)$. Hence,
the user observes an incomplete history $(T', <_e)$ called the \textit{observed
  history}. A transaction $t$ is compatible with a transaction $t'$ if and only
if, for each operation $o$ in $t$, there exists a unique operation $o'$ in $t'$
such that one of the following conditions holds: (1) $o = o'$; (2) $o =
  r_\theta(x, x_i)$ and $o' = r_\theta(x, \_)$; (3) $o = w_\theta(x, x_i, x_j)$
and $o' = w_\theta(x, \_ , x_j)$. A complete history $(T, <_e, <_v)$ is
compatible with an observed history $(T', <_e')$ if and only if $<_e = <_e'$
and for each $t \in T$, there exists a unique $t' \in T'$ such that $t$ is
compatible with $t'$.

\noindent{\bf Aborted and Intermediate Reads.}
There are two types of read operations that may cause anomalies: \emph{aborted reads} and \emph{intermediate reads}. An aborted read occurs when a committed transaction $t_j$ reads an version updated by an aborted transaction $t_i$. An intermediate read occurs when a committed transaction $t_j$ reads a version of an object updated by $t_i$, but not the last version created by $t_i$.

\noindent{\bf Item Dependencies between Transactions.}
Transactions that operate on the same object can form \emph{item dependencies}, primarily including three types of dependencies: \emph{item read-dependency}, \emph{item write-dependency} and \emph{item anti-dependency}.

\begin{definition}\label{def:item-dependency}
  For two committed transactions $t$ and $t'$ and a version $x_i$ of an object $x$,

  (1) $t'$ directly item read-depends on $t$ if $t$ installs $x_i$, and $t'$ reads $x_i$, denoted as $t \sxrightarrow{wr} t'$.

  (2) $t'$ directly item write-depends on $t$ if $t$ installs $x_i$, and $t'$ installs the version $x_{i + 1}$ of $x$ after $x_i$, denoted as $t \sxrightarrow{ww} t'$. If the version installed by $t'$ is newer than $x_i$ but is not the version immediately after $x_i$, $t'$ indirectly item write-depends on $t$, denoted as $t \sxtwoheadrightarrow{ww} t'$.

  (3) $t'$ directly item anti-depends on $t$ if $t$ reads $x_i$, and $t'$ installs the next version $x_{i + 1}$ of $x$ after $x_i$, denoted as $t \sxrightarrow{rw} t'$. If the version installed by $t'$ is newer than $x_i$ but is not the version immediately after $x_i$, $t'$ indirectly item anti-depends on $t$, denoted as $t \sxtwoheadrightarrow{rw} t'$.
\end{definition}

For ease of presentation, let $\mathbb{R}_x$, $\mathbb{W}_x$ and $\mathbb{A}_x$
represent the sets of direct item read-dependencies, item write-dependencies
and item anti-dependencies between the transactions in $T$ operating on a
certain object $x$, respectively. Let $\mathbb{R} = \bigcup_x \mathbb{R}_x$,
$\mathbb{W} = \bigcup_x \mathbb{W}_x$ and $\mathbb{A} = \bigcup_x
  \mathbb{A}_x$, where $x$ is an arbitrary object in the database. Let
$\mathbb{W}^+_x$ and $\mathbb{A}^+_x$ represent the sets of item
write-dependencies and item anti-dependencies (including both direct and
indirect ones) between the transactions in $T$ operating on a certain object
$x$, respectively. Finally, let $\mathbb{W}^+ = \bigcup_x \mathbb{W}^+_x$ and
$\mathbb{A}^+ = \bigcup_x \mathbb{A}^+_x$.

\noindent{\bf Predicate Dependencies between Transactions.}
Transactions that involve predicate reads and predicate writes may form \emph{predicate dependencies}. We define two types of predicate dependencies: \emph{predicate read-dependency} and \emph{predicate anti-dependency}.

\begin{definition}\label{def:predicate-dependency}
  For two committed transactions $t$ and $t'$ and a predicate $\theta$ valid for an object $x$,

  (1) $t'$ directly $\theta$-read-depends on $t$ if $t'$ $\theta$-reads a version $x_j$ of $x$, and $t$ installs the latest version $x_i$ of $x$ such that $x_i \le_v x_j$, $\theta(x_i) = \theta(x_j)$, and $\theta(x_i) \oplus \theta(x_{i - 1}) = \text{true}$ if $i > 1$, where $x_i \le_v x_j$ if and only if $x_i <_v x_j$ or $x_i = x_j$, and $\oplus$ is the exclusive or (XOR) operation. This dependency is denoted as $t \sxrightarrow{\theta, wr} t'$.

  (2) $t'$ directly $\theta$-anti-depends on $t$ if $t$ $\theta$-reads a version $x_i$ of $x$, and $t'$ installs the earliest version $x_j$ of $x$ such that $x_i <_v x_j$, $\theta(x_{j - 1}) = \theta(x_i)$, and $\theta(x_j) \oplus \theta(x_{j - 1}) = \text{true}$ if $j > 1$.  This dependency is denoted as $t \sxrightarrow{\theta, rw} t'$.
\end{definition}

Since a predicate write operation $w_\theta(x, x_i, x_j)$ can be considered as
applying $r_\theta(x, x_i)$ and $w(x, x_j)$ successively, we need not to define
``predicate write dependencies'' redundantly.

\noindent{\bf Serializability.}
Serializability requires that the execution of a set of transactions is equivalent to some serial execution, which can be characterized by Adya's formal definition~\cite{adya_generalized_2000} rephrased as follows.

\begin{definition}
  Given a set $T = \{t_1, t_2, \dots, t_n\}$ of committed transactions, the item dependencies and the predicate dependencies between the transactions in $T$ can be represented by a \emph{direct serialization graph (DSG)}, where each vertex represents a distinct transaction in $T$, and there is a directed edge $(t_i, t_j)$ from transaction $t_i$ to transaction $t_j$ if $t_i$ (item or predicate) depends on $ t_j$.
\end{definition}

\begin{theorem}\label{thm:Adya}
  A complete history of a set of transactions is serializable if and only if its DSG has no cycles, and there are neither aborted reads nor intermediate reads. 
\end{theorem}

To determine whether an observed history is serializable, we can test if there
is a compatible complete history that can serialize the observed history. This
condition is formally stated as follows.

\begin{theorem}\label{thm:serializable-observed-history}
  An observed history is serializable if and only if there exists a compatible complete history that is serializable.
\end{theorem}
\section{Black-box Serializability Verification}
\label{sec:black-box-serializability-verification}

\subsection{Problem Statement}
\label{sub:problem-statement}

\begin{figure}[htbp]
    \centering
    \includegraphics[width=0.8\linewidth]{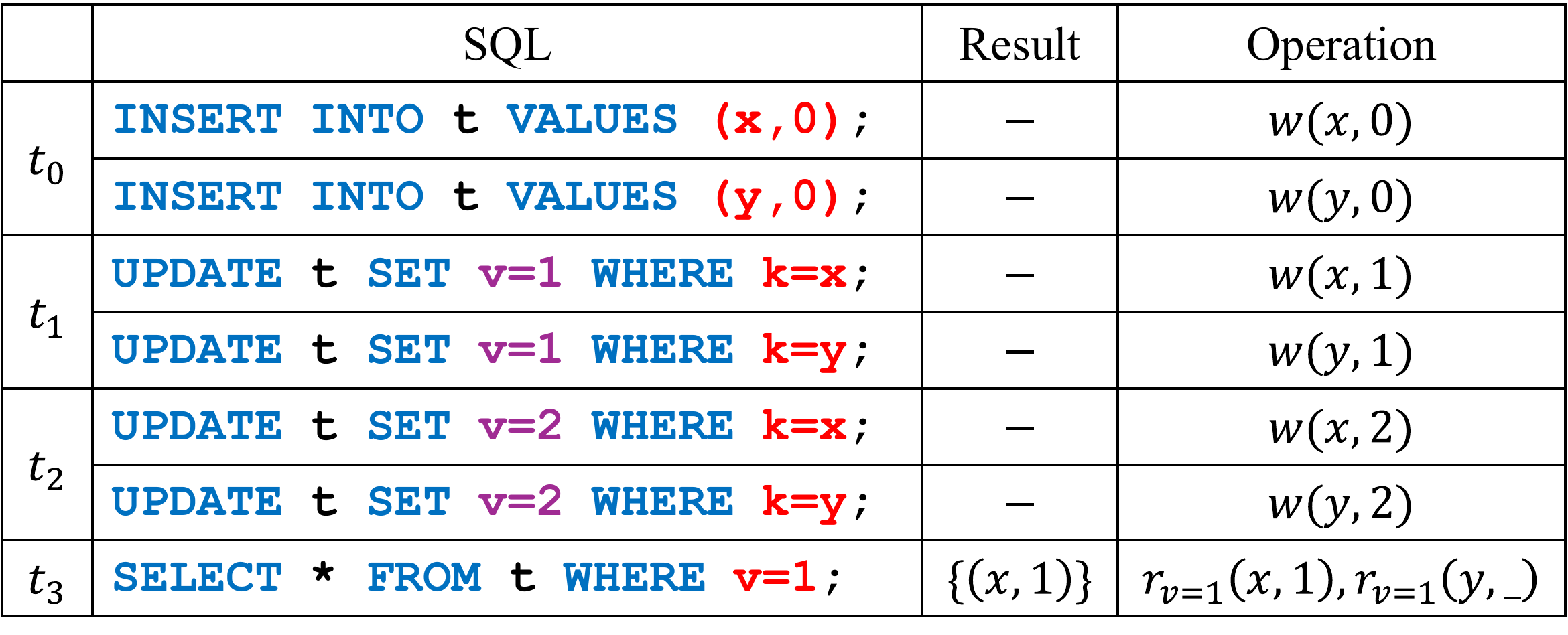}
    \caption{An observed history.}
    \Description[]{}
    \label{fig:history}
\end{figure}

Let's start with an example to illustrate black-box serializability verification. Consider the observed history of four transactions executed by four clients in a relational database, as shown in Figure~\ref{fig:history}. All SQL statements operate on a single relation $(k, v)$, where $k$ is the primary key. We use $k$ to refer to the object and $v$ to specify a version.

In transaction $t_3$, the only SQL statement is a predicate read based on $v = 1$. To determine the corresponding operations, we examine all objects in the history, namely $x$ and $y$. The result set contains only $x$ as $(x,1)$, indicating that one corresponding operation is $r_{v=1}(x, 1)$. The other operation, $r_{v=1}(y, \_)$, does not actually read any version of $y$, implying the version of $y$ scanned by the statement evaluated to be $F$ under the predicate $v=1$.

According to Theorem~\ref{thm:serializable-observed-history}, verifying serializability requires checking if a compatible complete history is serializable. The naive approach exhaustively checks all possible complete histories and verify if at least one is serializable. Constructing a compatible complete history involves determining the version order for versions ${0, 1, 2}$ of $x$ and $y$ and selecting a version of $y$ to complete $r_{v=1}(y, \_ )$. There are 6 possible version orders for both $x$ and $y$, and 2 candidates for $y$ (excluding version 1, since it must evaluate to false for $v = 1$). This results in $6 \times 6 \times 2 = 72$ complete histories. 

\begin{figure}[t]
    \centering
    \subfigure[Incomplete DSG]{
        \includegraphics[width=0.35\linewidth]{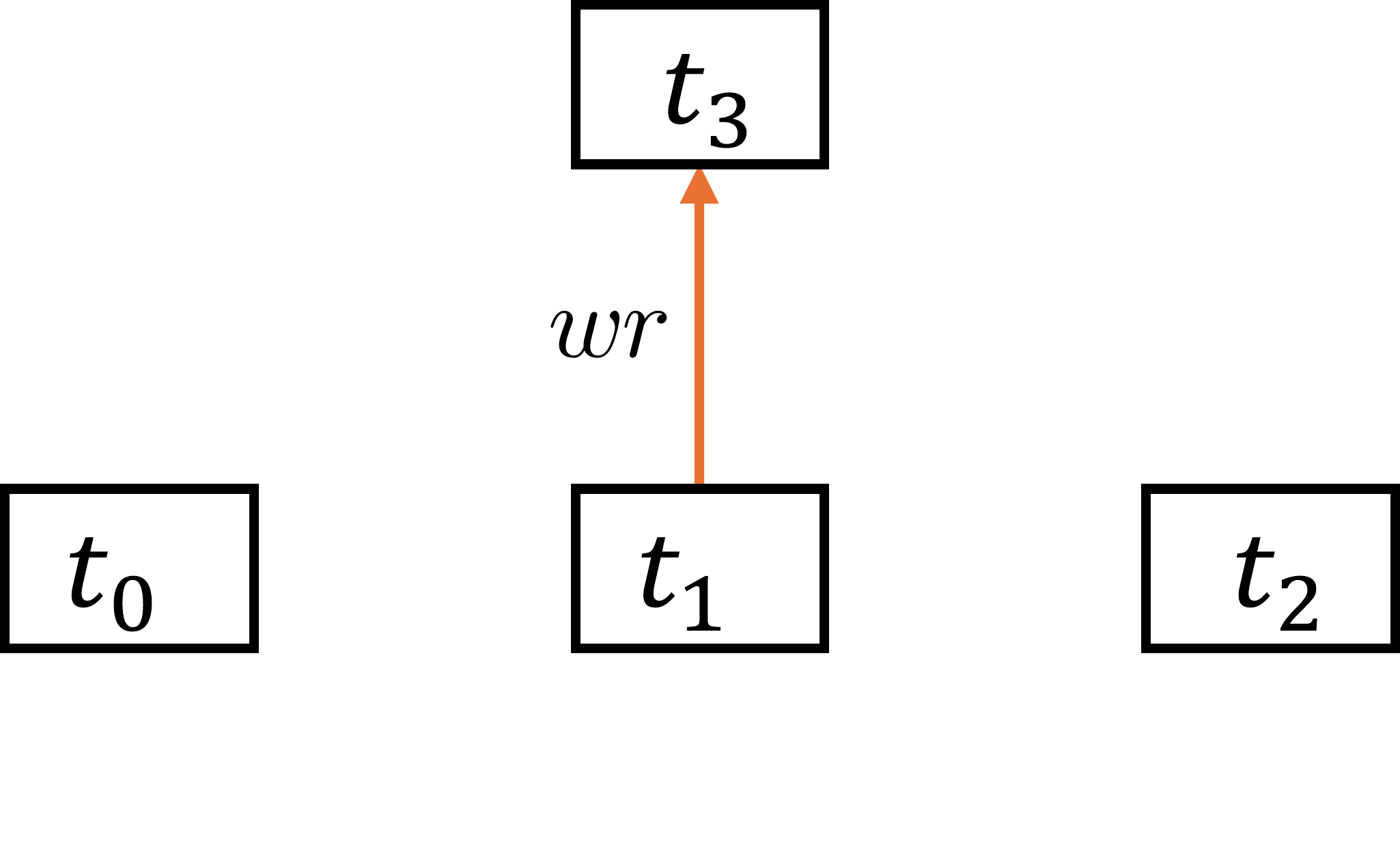}
        \label{fig:origin}
    }%
    \subfigure[DSG for $0,1,2$, $r_{v=1}(y, 2)$ ]{
        \includegraphics[width=0.35\linewidth]{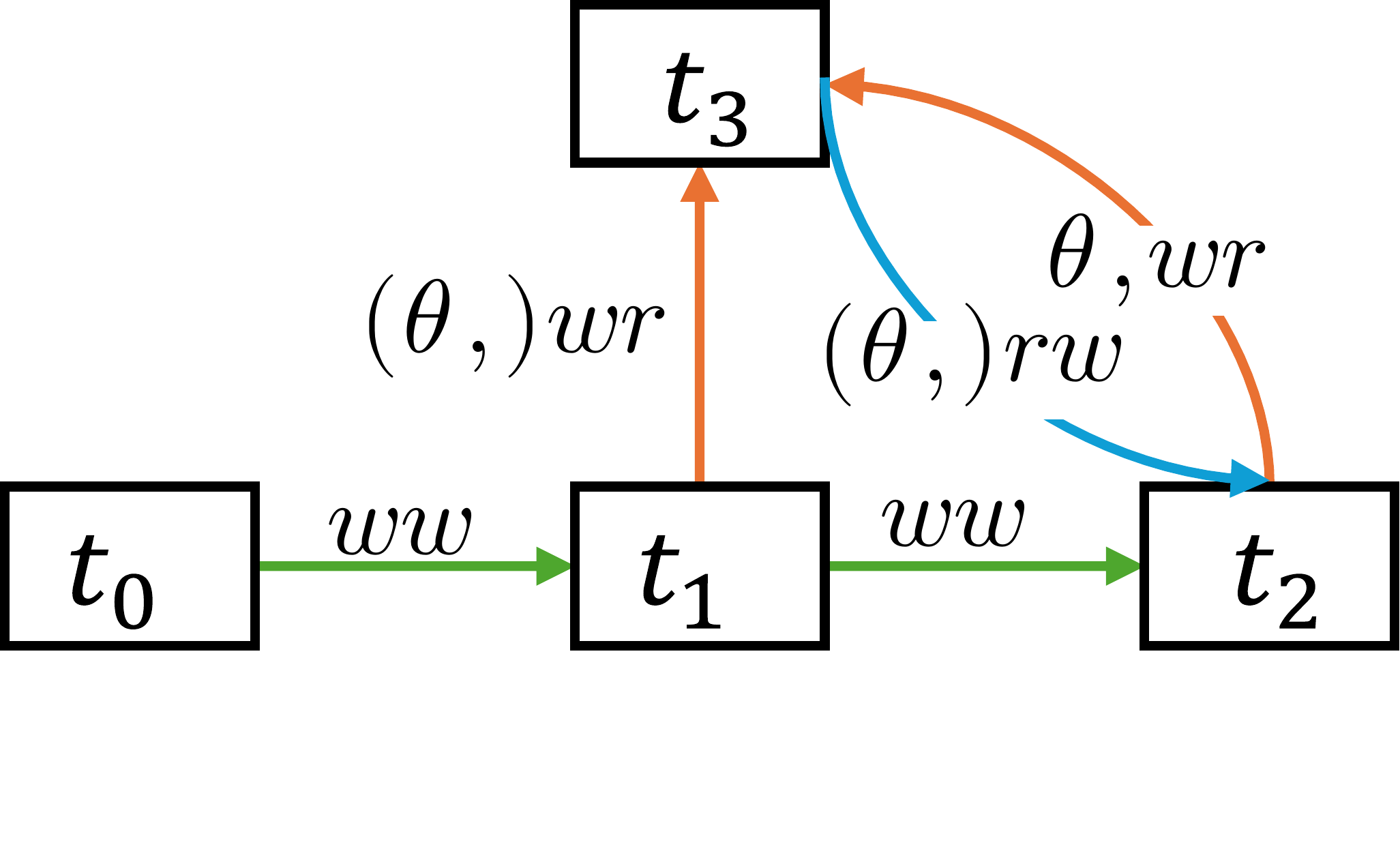}
        \label{fig:alpha}
    }%
    \newline
    \subfigure[DSG for $0,2,1$, $r_{v=1}(y, 2)$]{
        \includegraphics[width=0.35\linewidth]{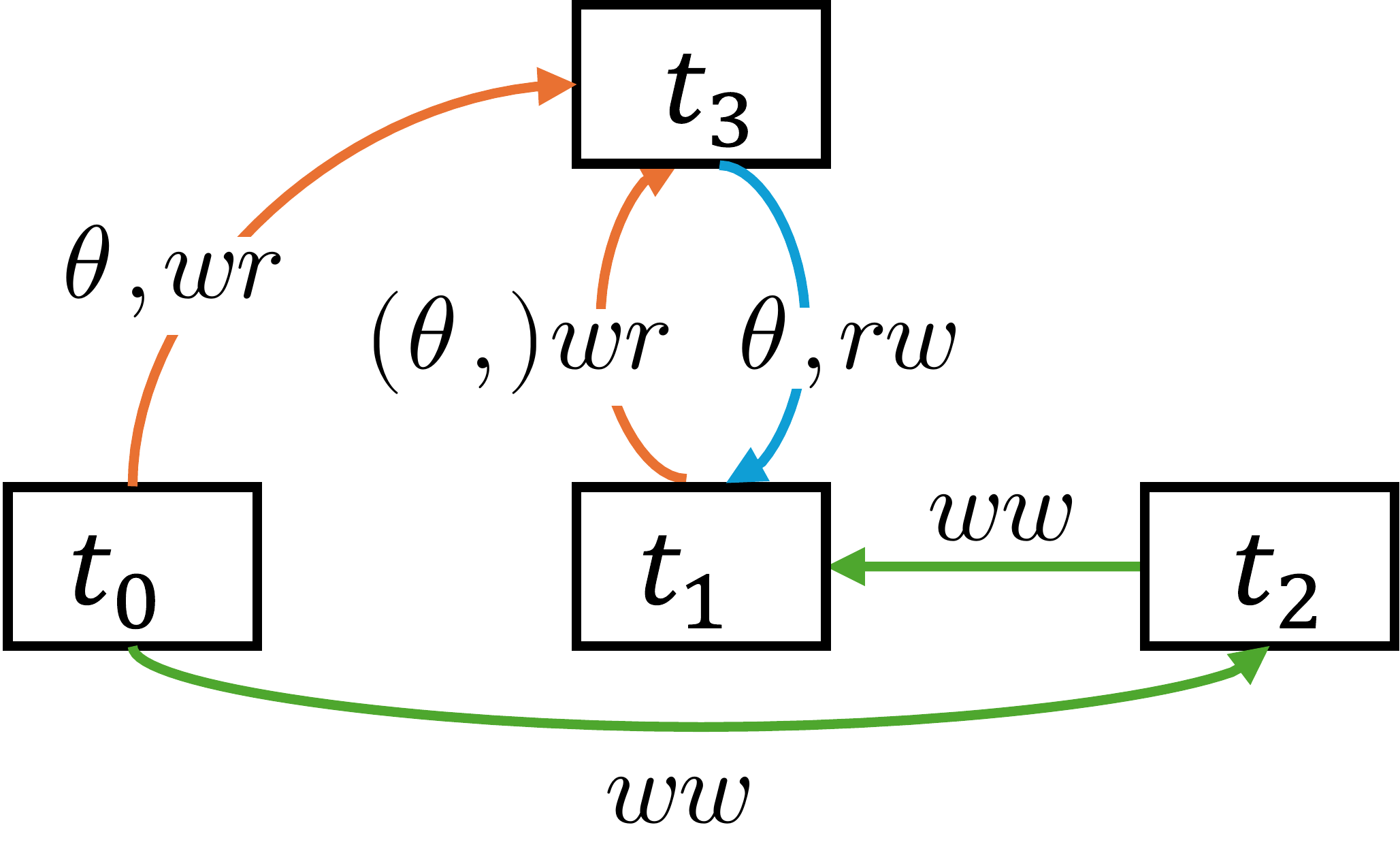}
        \label{fig:beta}
    }%
    \subfigure[DSG for $0,1,2$, $r_{v=1}(y, 1)$]{
        \includegraphics[width=0.35\linewidth]{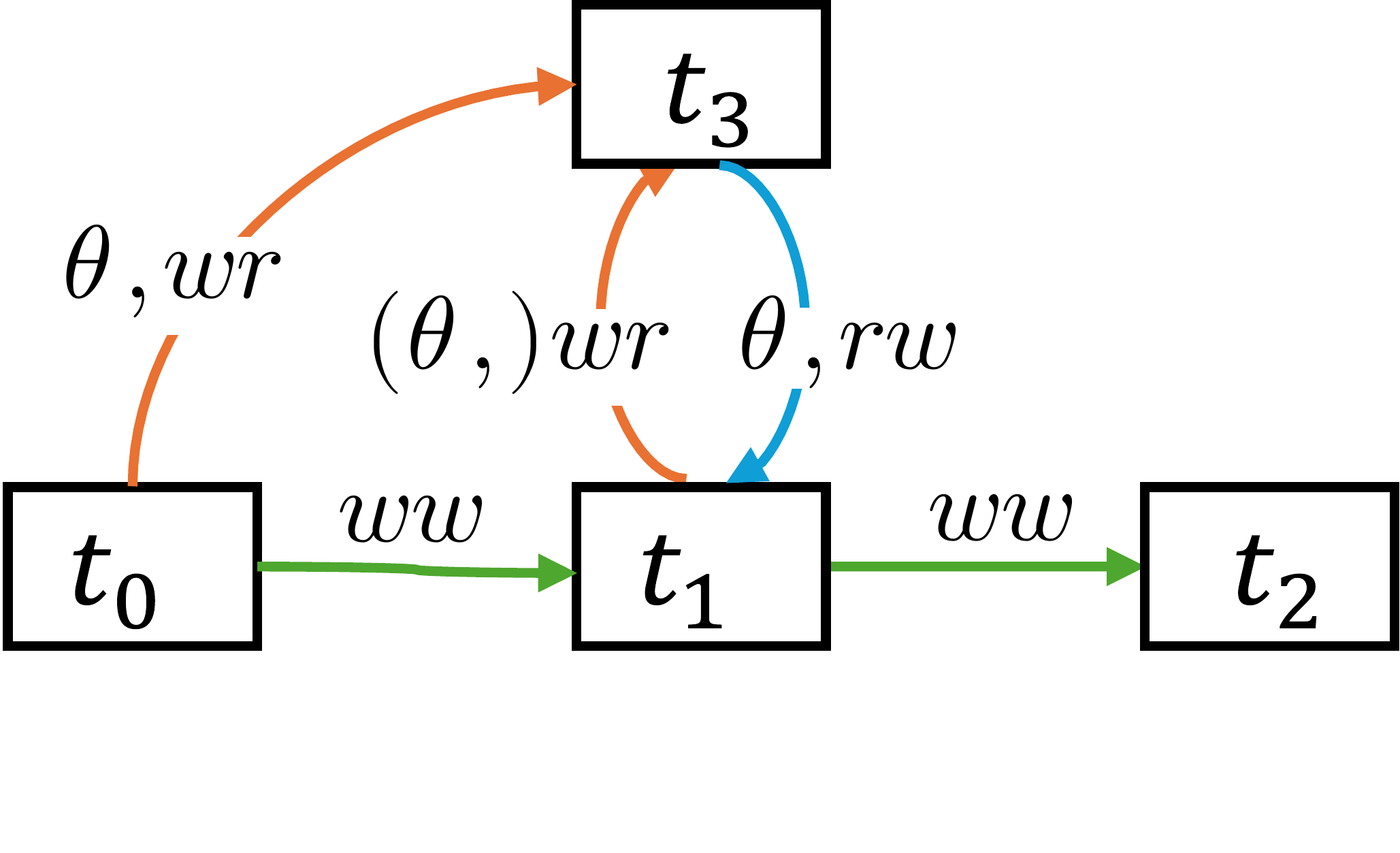}
        \label{fig:gamma}
    }%
    \vspace{-1em}
    \caption{Black-box serializability verification.}
    \Description[]{}
    \label{fig:cons}
    \vspace{-2em}
\end{figure}

For example, selecting version order $0 <_v 1 <v_2$ for $x$ and $y$ and choosing version $2$ of $y$ for $r{v=1}(y, \_)$ results in the DSG in Figure~\ref{fig:alpha}. Since the DSG contains a cycle, this complete history is not serializable by Theorem~\ref{thm:Adya}. Changing the version order to $0 <_v 2 <_v 1$ results in the DSG in Figure~\ref{fig:beta}, which also contains a cycle. In fact, after checking all possible complete histories, none of them are serializable, so the observed history is not serializable.

The black-box serializability verification problem is formalized as follows: Given an observed history $(T, <_e)$, determine if there exists a version order $<_v$ and a completion for all operations containing unknown versions such that the DSG of the complete history $(T, <_e, <_v)$ is acyclic and contains neither aborted reads nor intermediate reads.

\subsection{\textsf{Cobra}: A Baseline Approach}
\label{sub:cobra}

\textsf{Cobra}~\cite{tan_cobra_2020} is a baseline approach that cannot handle predicate reads and writes. Our approach extends \textsf{Cobra} (\S\ref{sec:our-method}).

\noindent{\bf Constraints.}
Given an observed history, the verification process starts from the \emph{known graph} $G = (V, E)$, where $V$ represents committed transactions, and $E$ represents direct item read-dependencies derived from the observation of the client. In order to add edges to $G$ that represents other dependencies, a set of constraints is established. Specifically, for committed transactions $t_i$ and $t_j$ installing versions $x_m$ and $x_n$ of $x$, either $x_m <_v x_n$ or $x_n <_v x_m$ holds in $<_v$, each case determining a set of dependency edges.

If $x_m <_v x_n$, the following set of edges can be added to $G$:
\begin{equation}\label{eq:item-constraint-edge-set-1}
    E_{ij} = \left\{ (t_i, t_j) \right\} \cup \left\{ (t_r, t_j) \mid t_i \xrightarrow[]{wr} t_r \in \mathbb{R}_x \right\}.
\end{equation}

Alternatively, if $x_n <_v x_m$, the following set of edges can be added to $G$:
\begin{equation}\label{eq:item-constraint-edge-set-2}
    E_{ji} = \left\{ (t_j, t_i) \right\} \cup \left\{ (t_r, t_i) \mid t_j \xrightarrow[]{wr} t_r \in \mathbb{R}_x \right\}.
\end{equation}

Since $E_{ij} \cap E_{ji} = \emptyset$ and $x_m <v x_n$ and $x_n <v x_m$ are exclusive, either $E{ij}$ or $E{ji}$ is in $G$. Hence, $(E_{ij}, E_{ji})$ forms a constraint. In Figure~\ref{fig:origin}, transactions $t_0, t_1, t_2$ install versions $0,1,2$ of $x$ and $y$, respectively. The constraint set is $\{(E_{01}, E_{10}),(E_{02}, E_{20}),(E_{12}, E_{21})\}$, where $E_{01} = \{ (t_0, t_1)\}$, $E_{10} = \{(t_1, t_0),(t_3,t_0) \}$, $E_{02} = \{ (t_0, t_2)\}$, $E_{20} = \{(t_2, t_0) \}$, $E_{12} = \{ (t_1, t_2),(t_3,t_2)\}$, $E_{21} = \{(t_2, t_1) \}$.

\noindent{\bf Compatible Graphs.}
Given the known graph $G$ and a set $C$ of constraints, a \emph{compatible graph} $G' = (V, E')$ can be constructed that satisfies the following condition: for all constraints $(E_{ij}, E_{ji}) \in C$, the following proposition is true:
\begin{equation}\label{eq:compatible-graph}
    \left( E_{ij} \subseteq E' \land E_{ji} \cap E' = \emptyset \right) \lor \left( E_{ij} \subseteq E' \land E_{ji} \cap E' = \emptyset \right).
\end{equation}
Hence, a compatible graph $G'$ can be obtained by adding either $E_{ij}$ or $E_{ji}$ to $G$ for each constraint $(E_{ij}, E_{ji}) \in C$. The choice for $E_{ij}$ or $E_{ji}$ for all constraints $(E_{ij}, E_{ji}) \in C$ is called an \emph{assignment}. The compatible graph $G'$ is normally a proper supergraph of the DSG. If $G'$ is acyclic, the DSG is also acyclic, and therefore, the history is serializable.

Consider the known graph $G$ in Figure~\ref{fig:origin} and the constraint set $\{(E_{01}, E_{10}), (E_{02}, E_{20}), (E_{12}, E_{21})\}$. Choosing $E_{01}, E_{02}, E_{12}$ yields the compatible graph in Figure~\ref{fig:alpha} (excluding $(t_2,t_3)$), while selecting $E_{01}, E_{02}, E_{21}$ results in Figure~\ref{fig:beta} (excluding $(t_0,t_3)$ and $(t_3,t_1)$). Both graphs are acyclic, making \textsf{Cobra} consider the history serializable. However, this conflicts with the verification in Section~\ref{sub:problem-statement}. The discrepancy arises because \textsf{Cobra} does not handle predicate reads and overlooks versions that clients do not actually read, such as $r_{v=1}(y, \_)$. These overlooked versions influence the dependencies between transactions. For $r_{v=1}(y, \_)$, there must be a transaction $t_i$ that installs a version $y_m$ of $y$ making $v=1$ false, and $t_1$ evaluates $v=1$ on it. Additionally, if another transaction $t_j$ installs a version after $y_m$ that makes $v=1$ true, then $t_j$ must anti-depend on $t_1$.

\noindent{\bf Satisfiability Problem Solving.}
To verify serializability, it suffices to construct an acyclic compatible graph $G'$, a nontrivial problem due to $2^{|C|}$ possible assignments. \textsf{Cobra} converts this to a SAT problem by introducing boolean variables $b_{ij}$ for transaction pairs $t_i, t_j$, setting $b_{ij} = T$ if $(t_i, t_j) \in G'$. Each constraint $(E_{ij}, E_{ji}) \in C$ is formulated as the clause $\left( P_{ij} \land \neg P_{ji} \right) \lor \left( \neg P_{ij} \land P_{ji} \right)$, where $P_{ij} = \bigwedge_{(t_m, t_n) \in E_{ij}} b_{mn}$, and $P_{ji} = \bigwedge_{(t_p, t_q) \in E_{ji}} b_{pq}$. Cobra tests if the following formula can be satisfied

\setlength{\abovedisplayskip}{5pt}
\setlength{\belowdisplayskip}{5pt}
\begin{equation}\label{eq:cobra-sat-formulation}
    \bigwedge_{(E_{ij}, E_{ji}) \in C} \left( P_{ij} \land \neg P_{ji} \right) \lor \left( \neg P_{ij} \land P_{ji} \right),
\end{equation}
subject to the constraint that the compatible graph $G'$ induced by the edges $(t_i, t_j)$ corresponding to $b_{ij} = T$ is acyclic. \textsf{Cobra} uses the MonoSAT solver~\cite{bayless_sat_2015} to solve the SAT problem.

\subsection{Limitations of \textsf{Cobra}}
\label{sub:cobra-limitations}

\noindent{\bf Limitation 1: Lack of Predicate Support.}  
\textsf{Cobra} lacks support for predicate reads and writes, limiting its applicability to real-world transactions.  

\noindent{\bf Limitation 2: Excessively Generates Constraints.}  
\textsf{Cobra} creates a constraint for every transaction pair operating on the same object, resulting in $|C| = \sum_{x \in X} (|T_x|^2-|T_x|)$ constraints, where $|T_x|$ denotes the number of transactions modifying object $x$.  

\noindent{\bf Limitation 3: Poor Scalability.}  
\textsf{Cobra} relies on Warshall's algorithm ($O(|T|^3)$ time complexity and $O(|T|^2)$ space complexity) to maintain transitive closure, making it impractical for large history.  

\noindent{\bf Limitation 4: Inefficient SAT Problem Formulation.}  
\textsf{Cobra} assigns a boolean variable to each constraint $(E_{ij}, E_{ji}) \in C$ and enforces mutual exclusivity through Equation~\ref{eq:cobra-sat-formulation}. However, within a constraint, the boolean variables for the edges within the same set are uniformly either all $T$ or all $F$.

\noindent{\bf Limitation 5: Unsuitable Solver.}  
\textsf{Cobra} uses MonoSAT for SAT solving but disregards precomputed transitive closure data, redundantly performing dynamic topological sorting for cycle detection.

\section{Our Method \textsf{Vbox}}
\label{sec:our-method}

Our black-box serializability verification method called \textsf{Vbox} is developed from \textsf{Cobra}. To address the limitations of \textsf{Cobra}, we propose a series of targeted technical improvements.

(1) To overcome Limitation~1 (\emph{lack of predicate support}), we introduce predicate constraints for predicate read-dependencies and anti-dependencies (\S\ref{sub:predicate-constraints}).

(2) To overcome Limitation~2 (\emph{excessive number of constraints}), we leverage client transaction timestamps for a more complete known graph, filtering unnecessary constraints (\S
\ref{sub:known-graph-construction}) and applying pruning techniques (\S
\ref{sub:constraint-reduction}).

(3) To overcome Limitation~3 (\emph{poor scalability}), we propose a compact transitive closure matrix. Based on it, we develop optimized graph algorithms to support fast edge insertion and edge deletion, which reduces both time and space complexities and significantly improves scalability given a large number of transactions (\S\ref{sub:efficient-reachability-test}).

(4) To overcome Limitation~4 (\emph{inefficient SAT problem formulation}), we simplify the SAT problem formulation by assigning a boolean variable for each item constraint. The simplified formulation avoids using a large number of formulas to characterize the mutual exclusion of the edge sets in the constraints (\S\ref{sub:simplified-sat-problem-formulation}).

(5) To overcome Limitation~5 (\emph{unsuitable solver}), we design a customized solver for the SAT problem formulated for serializable verification. The solver reuses the data structures built during problem formulation to enable efficient cycle detection and leverages the transactions' timestamps to guide the search (\S\ref{sub:customized-sat-problem-solver}).

\subsection{Handling Predicate Operations}
\label{sub:predicate-constraints}
To verify serializability of an observed history $(T, <_e)$, we check for a version order $<_v$ and completion of unknown $\theta$-reads such that $(T, <_e, <_v)$ is serializable. \textsf{Cobra} can identify a ``valid'' version order but ignores predicate operations. When predicates are considered, the version order found by \textsf{Cobra} may become invalid. In Figure~\ref{fig:history}, \textsf{Cobra} may establish $0 <_v 1 <_v 2$ for both $x$ and $y$, but completing $r_{v=1}(y, \_)$ with version $0$ or $2$ results in a non-serializable history (Figures~\ref{fig:beta} and \ref{fig:gamma}). This is because the additional predicate dependencies introduced by $r_{v=1}(y, \_)$ alter the DSG connectivity and create cycles.

To complete the verification, one approach is to use \textsf{Cobra} to identify candidate version orders and then validate them with predicate operations. However, \textsf{Cobra} is designed to find a single valid version order, not to generate multiple candidates. To address this, we introduce \emph{predicate constraints} to formalize predicate dependencies, distinguishing them from the \emph{item constraints} used by \textsf{Cobra}. We verify the serializability by uniformly making assignments for item constraints and predicate constraints.

\noindent{\bf Validating version order.} We validate version orders by first constructing an incomplete DSG with only item dependencies, then adding predicate dependencies and checking for cycles.

Predicate dependencies from known $\theta$-reads do not affect DSG connectivity. In Figure~\ref{fig:history}, the known $\theta$-read $r_{v=1}(x, 1)$ in $t_3$ introduces an item dependency $(t_1, t_3)$ and a predicate dependency $(t_0, t_3)$. Removing $(t_0, t_3)$ does not impact connectivity since $t_0 \to t_2 \to t_1 \to t_3$ remains via item dependencies. This path is guaranteed because the transaction $t_i$ that $t_3$'s predicate read depends on must install a version before or equal to the version $t_3$ actually reads. This ensures a path from $t_i$ to $t_3$ through item write and read dependencies. Similarly, predicate anti-dependencies for known $\theta$-reads do not change connectivity. Thus, predicate dependencies for known $\theta$-reads need not be considered when validating the version order.

However, predicate dependencies for unknown $\theta$-reads do affect DSG connectivity. In Figure~\ref{fig:alpha}, the predicate read dependency for $r_{v=1}(y, \_)$ is $(t_2, t_3)$, and removing this edge eliminates the cycle in the DSG. Each unknown $\theta$-read $r_\theta(y, \_)$ completes itself by selecting a version $y_i$ satisfying $\theta(y_i) = \text{false}$. Each selected $y_i$ generates an edge set containing predicate write and anti-dependency edges, denoted as $E_{\theta,y_i}$. All such edge sets form a \emph{validation constraint} $\{E_{\theta,y_i} \mid \theta(y_i) = \text{false}\}$. For the history in Figure~\ref{fig:history}, if the version order is $0 <_v 2 <_v 1$ for $y$, the validation constraint for $r_{v=1}(y, \_)$ is $\{E_{v=1, 0}, E_{v=1, 2}\}$, where $E_{v=1, 0} = \{ (t_0, t_3), (t_3, t_1) \}$ and $E_{v=1, 2} = \{(t_0, t_3), (t_3, t_1)\}$.

Selecting an edge set from each validation constraint and adding it to the incomplete DSG constructs a compatible graph. If the resulting graph is acyclic, the version order is valid, and the history is serializable.

\noindent{\bf Predicate constraints.} 
Validation constraints do not resolve the issue that \textsf{Cobra} cannot generate version order candidates. If we can uniformly assign both item and validation constraints, and resulting compatible graph is acyclic, the history is serializable. However, the edges in the validation constraint remain undetermined when the version order is unknown, which restricts us to assign the validation constraint only after the item constraints are assigned. To address this, we introduce the \textit{predicate constraint}, which allows us to determine the predicate dependency edges once part of the version order is known.

For an unknown $\theta$-read $r_\theta(y,\_)$ in transaction $t$, the transaction $t_m$ that $t$ predicate read depends on must install a version $y_i$ of $y$ such that $\theta(y_i) = \text{false}$. Additionally, the transaction $t_n$ that predicate anti-depends on $t$ must install a version $y_j$ such that $\theta(y_j) = \text{true}$ and $y_i <_v y_j$, implying that $t_m \sxtwoheadrightarrow{ww} t_n$. 
Let $F_{\theta,y}$ denote the set of transactions that install a version $y_i$ where $\theta(y_i) = \text{false}$. For each $t_m \in F_{\theta,y}$, we construct the following edge set:

\begin{equation}\label{eq:predicate-constraint}
    E_{\theta,t_m} = \left\{(t_m, t)\right\} \cup \left\{(t, t_n) \mid t_m \sxtwoheadrightarrow{ww} t_n, t_n \text{ installs } y_j, \theta(y_j) = \text{true} \right\}
\end{equation}
The first part of the edge set is a possible predicate read dependency and the other part are predicate anti-dependencies.
All such edge sets form a predicate constraint, formally defined as $\{ E_{\theta,t_m} \mid t_m \in F_{\theta,y} \}$. 


It is important to note that the predicate read dependency $(t_m,t)$ in $E_{\theta,t_m}$ may not strictly follow the Definition~\ref{def:predicate-dependency}, as we do not restrict the version $y_i$ installed by $t_m$ to satisfy $ \theta(y_i) \oplus \theta(y_{i-1}) = \text{false}$, where $y_{i-1}$ is the direct processor of $y_i$ according to $<_v$. Second, $E_{\theta,t_m}$ may contain more than one predicate anti-dependency edge $(t,t_n)$. This is because we do not restrict the version $y_j$ installed by $t_n$ to be the earliest version such that $\theta(y_j) = \text{true}$ and $y_i <_v y_j$. 

Finding the earliest version or predecessor of a version can only be done after the complete version order is established, which prevents the unified assignment of predicate and item constraints. \autoref{thm:constraint-serializable} demonstrates the correctness of both predicate and item constraints.

\begin{theorem}\label{thm:constraint-serializable}
An observed history is serializable if and only if the history contains no intermediate or aborted read anomalies and there exists an assignment for its item constraint set and predicate constraint set that makes the generated compatible graph acyclic.
\end{theorem}

\noindent{\bf Unified assignment.} 
An edge $(t, t_n)$ is included in $E_{\theta,t_m}$ only if $t_m \sxtwoheadrightarrow{ww} t_n$. To unify item and predicate constraint assignments, we first add all potential edges $(t, t_n)$ where $\theta(y_j) = \text{true}$ to $E_{\theta,t_m}$, marking them as undetermined. Once the item constraint $(E_{mn}, E_{nm})$ is assigned and $E_{mn}$ is chosen (implying $t_m \sxtwoheadrightarrow{ww} t_n$), $(t, t_n)$ is determined. Predicate constraint assignment involves selecting an edge set and adding the determined edges to the known graph.

In the history of Figure~\ref{fig:history}, the item constraint set we construct is $\{(E_{01}, E_{10}), (E_{02}, E_{20}), (E_{12}, E_{21})\}$, and the initial predicate constraint set is $\{E_{v=1,t_0}, E_{v=1,t_2}\}$, where $E_{v=1,t_0} = \{(t_0,t_3), (t_3,t_1)\}$ and $E_{v=1,t_2} = \{(t_2,t_3), (t_3,t_1)\}$, with $(t_3,t_1)$ marked as undetermined. Assigning $(E_{01}, E_{10})$ and choosing $E_{01}$ makes $(t_3,t_1)$ in $E_{v=1,t_0}$ determined. Further choosing $E_{v=1,t_0}$ for the predicate constraint and $E_{02}$, $E_{21}$ for item constraints yields the DSG in Figure~\ref{fig:beta}.

This unified assignment eliminates the need to generate version order candidates and offers two additional benefits. First, it reduces the search space by determining edges in advance and pruning invalid edge sets (\S\ref{sub:constraint-reduction}). Second, it simplifies problem formulation (\S\ref{sub:simplified-sat-problem-formulation}). When formalizing $(t, t_n)$ in $E_{\theta,t_m}$, we focus solely on the write dependency between $t_m$ and $t_n$, without considering other transactions installing versions of the same object.

\subsection{Effective Construction of Known Graph}
\label{sub:known-graph-construction}

Our next design goal is to construct a more complete known graph by identifying more dependencies based on the observation from the client and avoid unnecessary constraints. To this end, we record the client timestamps of the transactions in the history. For any transaction $t$ in the history, we record the \emph{client start timestamp} $\text{s}(t)$ of $t$ when the client sends $t$'s \texttt{begin} command to the server, and we record the \emph{client end timestamp} $\text{e}(t)$ of $t$ when the client receives the response of $t$'s \texttt{commit} or \texttt{abort} command. These timestamps are based on the wall-clock time provided by the client machine and need to be synchronized between different clients. Due to the transmission latency, $\text{s}(t)$ is earlier than the actual start timestamp $\text{s}^r(t)$ of $t$ on the server, and $\text{e}(t)$ is later than the actual end timestamp $\text{e}^r(t)$ of $t$ on the server.

\noindent{\bf Time-Dependencies.}
To obtain a more complete known graph $G$, we define dependencies between transactions based on their start and end timestamps. For two committed transactions $t_i$ and $t_j$, if $\text{e}(t_i) \leq \text{s}(t_j)$,  it indicates that the DBMS has completed and committed all operations of transaction $t_i$ before receiving the \texttt{begin}  operation of $t_j$. In this case, $t_j$ is said to \emph{time-depend} on $t_i$. Let $\mathbb{T}$ denote the set of all time-dependencies between the transactions in the history. 

Some item write-dependencies and item anti-dependencies can be derived from time-dependencies. If two committed transactions $t_i$ and $t_j$ install different versions of the same object, and $t_j$ time-depends on $t_i$, we have $\text{e}^r(t_i) \leq \text{s}^r(t_j)$, so $t_j$ item write-depends on $t_i$, that is, $t_i \sxrightarrow{ww} t_j$ or $t_i \sxtwoheadrightarrow{ww} t_j$. Let $\mathbb{W}^{+t} \subseteq \mathbb{W}^+$ denote the set of item write-dependencies inferred from the time-dependencies in $\mathbb{T}$.

If a committed transaction $t_i$ reads an object $x$, another committed transaction $t_j$ installs $x$, and $t_j$ time-depends on $t_i$, we have that $t_j$ item anti-depends on $t_i$, that is, $t_i \sxrightarrow{rw} t_j$ or $t_i \sxtwoheadrightarrow{rw} t_j$. Let $\mathbb{A}^{+t} \subseteq \mathbb{A}^+$ denote the set of item anti-dependencies inferred from the time-dependencies in $\mathbb{T}$.

\noindent{\bf Constructing Known Graph.}
Note that an item anti-dependency can be derived from an item read-dependency followed by an item write-dependency. If a committed transaction $t_i$ read-depends on a committed transaction $t_r$ with respect to an object $x$, and another committed transaction $t_j$ item write-depends on $t_i$ also with respect to $x$, we have that $t_j$ item anti-depends on $t_r$ with respect to $x$, that is, $t_r \sxrightarrow{ww} t_j$ or $t_r \sxtwoheadrightarrow{ww} t_j$. Let $\mathbb{A}^{+w} \subseteq \mathbb{A}^+$ denote the set of item anti-dependencies inferred from the item read-dependencies in $\mathbb{R}$ and the item write-dependencies in $\mathbb{W}^{+t}$.

After identifying the item read-dependencies in $\mathbb{R}$, the time-dependencies in $\mathbb{T}$, the item write-dependencies in $\mathbb{W}^{+t}$, and the item anti-dependencies in $\mathbb{A}^{+t} \cup \mathbb{A}^{+w}$ from the client, we can construct the known graph $G$ by adding edges representing all these identified dependencies.

Different from item read-dependency, item write-dependency and item anti-dependency, time-dependency is not defined based on read/write operations on objects. Theorem~\ref{thm:time-order} ensures the correctness of adding time-dependencies to the known graph $G$, and we prove this theorem in the appendix.

\begin{theorem}\label{thm:time-order}
If the concurrency control protocol adopted by the DBMS is Serializable Snapshot Isolation (SSI)~\cite{CahillRF09}, Two-Phase Locking (2PL)~\cite{RosenkrantzSL13a}, Optimistic Concurrency Control (OCC)~\cite{KungR81}, Timestamp Ordering (TO)~\cite{BernsteinG81}, or Percolator~\cite{PengD10}, adding time-dependency edges to a DSG that is compatible with the observed history does not change the reachability states between the vertices in the DSG.
\end{theorem}

\subsection{Constraint Reduction}
\label{sub:constraint-reduction}

The number of constraints determines the scale of the SAT problem to be solved and further the efficiency of verification. We propose a series of methods to reduce the number of constraints, including avoiding and pruning of useless constraints.

\noindent{\bf Item Constraint Avoidance.}
In \textsf{Cobra}, an item constraint $(E_{ij}, E_{ji})$ is constructed for each pair of transactions $t_i$ and $t_j$ that install different versions of the same object (see Eq.~\eqref{eq:item-constraint-edge-set-1} and Eq.~\eqref{eq:item-constraint-edge-set-2}). In fact, it is unnecessary to construct this constraint if $t_i$ and $t_j$ do not overlap in time, that is, $\text{e}(t_i) \leq \text{s}(t_j)$ or $\text{e}(t_j) \leq \text{s}(t_i)$. This is because the item write-dependency between $t_i$ and $t_j$ has already been inferred from this time-dependency and has been added to $\mathbb{W}^{+t}$, and the item anti-dependencies inferred from this item write-dependency have already been added to $\mathbb{A}^{+w}$.

\noindent{\bf Item Constraint Pruning.}
Let $(E_{ij}, E_{ji})$ be an item constraint. For any edge $(u, v) \in E_{ij}$ (or $(u, v) \in E_{ji}$), if there exists a directed path from $v$ to $u$ in the known graph $G$, adding the directed edge $(u, v)$ to $G$ would form a cycle in the compatible graph $G'$, indicating that such history is not serializable. Therefore, none of the edges in $E_{ij}$ (or $E_{ji}$) can be added to $G$, and only the edges in $E_{ji}$ (or $E_{ij}$) can be added to $G$. For this reason, we can remove this constraint and add all the edges in $E_{ji}$ (or $E_{ij}$) to $G$. 

\noindent{\bf Predicate Constraint Avoidance.}
For each $\theta$-read $r_\theta(y, \_)$, we construct a predicate constraint $c = \{ E_{\theta,t_m} \mid t_m \in F_{\theta,y} \}$. For an edge $(t_i, t_j) \in E_{\theta,t_m}$, if $\text{e}(t_i) \leq \text{s}(t_j)$, then $(t_i, t_j)$ has already been added to the known graph as a time dependency edge. On the other hand, if $\text{e}(t_j) \leq \text{s}(t_i)$, adding $(t_i, t_j)$ would form a cycle with the existing time dependency edge $(t_j, t_i)$. Any edge $(t_i, t_j) \in E_{\theta,t_m}$ where $t_i$ and $t_j$ do not overlap in time is unnecessary and should be excluded from $E_{\theta,t_m}$. After removing such edges, $E_{\theta,t_m}$ may become empty or identical to an existing edge set in $c$. Therefore, we retain only the non-empty, distinct edge sets in the predicate constraint $c$.

\noindent{\bf Predicate Constraint Pruning.}
Let $\{ E_{\theta,t_m} \mid t_m \in F_{\theta,y} \}$ be a predicate constraint. Initially, for each edge set $E_{\theta,t_m}$, only the predicate read dependency edge $(t_m, t)$ is determined, while predicate anti-dependency edges $(t, t_n)$ remain undetermined.

For an undetermined edge $(t, t_n)$ derived from a write dependency edge $(t_m, t_n)$, if there is a directed path from $t_m$ to $t_n$ in the known graph $G$, then $(t, t_n)$ becomes determined because $(t_m, t_n)$ has already been or will be added to $G$ during item constraint pruning. If there is a directed path from $t_n$ to $t_m$ in $G$, we remove $(t, t_n)$ from $E_{\theta,t_m}$, as $(t_m, t_n)$ will never be added to $G$.

For a determined edge $(t_i, t_j) \in E_{\theta,t_m}$, if there is a directed path from $t_j$ to $t_i$ in $G$, adding this edge to $G$ would form a cycle, so we remove $E_{\theta,t_m}$ from the predicate constraint. If only one edge set remains in the predicate constraint, we add the determined edges to $G$. Note that we cannot remove the constraint immediately because there may be undetermined edges in the unique edge set of the constraint. Once all these undetermined edges become determined during pruning, we add them to $G$ and then remove the constraint.


\subsection{Efficient Reachability Test}
\label{sub:efficient-reachability-test}

Constraint pruning and our SAT problem solving method (\S\ref{sub:customized-sat-problem-solver}) frequently ask if two vertices (transactions) are reachable in the known graph $G$. 
For efficient reachability test, we maintain the transitive closure of $G$, a matrix $R$ where the element $R_{ij}$ indicates whether transaction $t_j$ is reachable from transaction $t_i$ in $G$. As we observed, it is inefficient to explicitly maintain the full transitive closure. By utilizing the start and the end timestamps of the transactions, we propose the \textit{compact transitive closure} of $G$. Here, we study how to efficiently compute and update the compact transitive closure.

\noindent{\bf Compact Transitive Closure.}
The compact transitive closure leverages the observation that for two committed transactions $ t_i $ and $ t_j $, if $ \text{e}(t_i) \leq \text{s}(t_j) $, the edge $ (t_i, t_j) $ is added to the graph $ G $, indicating $ t_j $ is reachable from $ t_i $. For transactions $ t_i $ and $ t_j $ that do not overlap ($ \text{e}(t_i) \leq \text{s}(t_j) $ or $ \text{e}(t_j) \leq \text{s}(t_i) $), their reachability can be determined directly from their timestamps without explicitly recording it in the transitive closure.

The design of the compact transitive closure is as follows. Transactions are first sorted by their start timestamps, resulting in the sorted list $ T^s $. For a transaction $ t_i $, let $ t_{l_i} $ and $ t_{r_i} $ denote the first and last transactions in $ T^s $ that overlap with $ t_i $, respectively. Since $ T^s $ is sorted, transactions before $ t_{l_i} $ or after $ t_{r_i} $ do not overlap with $ t_i $, and their reachability can be inferred from timestamps. In the compact transitive closure, we only track reachability between $ t_i $ and transactions $ t_j $ where $ t_{l_i} \leq t_j \leq t_{r_i} $.

Instead of using a hash map, we design an array-based data structure for the compact transitive closure, which is more cache-efficient. The structure consists of the main array $ R $ and two auxiliary arrays $ L $ and $ O $. For a transaction $ t_i $ and each transaction $ t_j $ between $ t_{l_i} $ and $ t_{r_i} $ in $ T^s $, the reachability status $ r(t_i, t_j) $ is stored in $ R $. The auxiliary arrays $ L $ and $ O $ efficiently locate these entries: $ L[i] $ stores the index $ l_i $ of the first overlapping transaction $ t_{l_i} $, and $ O[i] $ stores the starting index in $ R $ for $ t_i $'s reachability records. Thus, $ r(t_i, t_j) $ for $ t_{l_i} \leq j \leq t_{r_i} $ is stored at $ R[O[i] + j - L[i]] $.

\begin{figure}[t]
    \centering
    \includegraphics[width=0.7\linewidth]{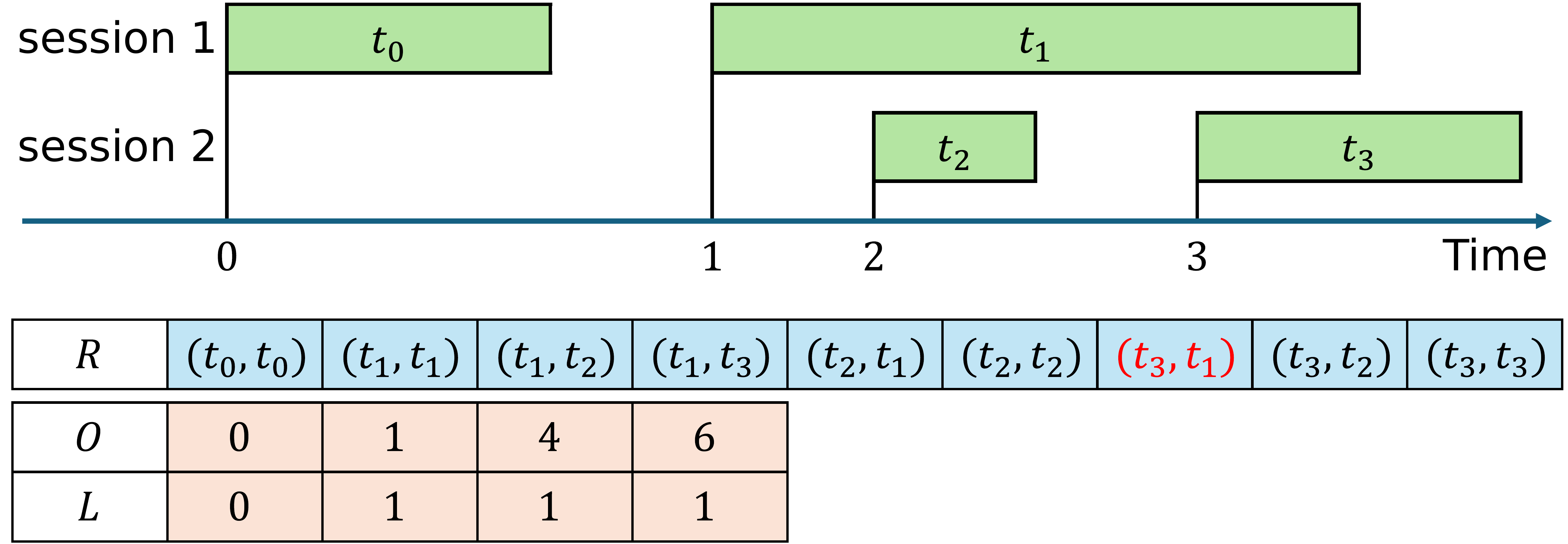}
    \caption{Compact transitive closure.}
    \label{fig:compact-transitive-closure}
    \vspace{-2em}
    \Description{}
\end{figure}

\autoref{fig:compact-transitive-closure} illustrates an example of compact transitive closure with four transactions $ t_0 $--$ t_3 $ sorted by their start timestamps. For $ t_0 $, since it does not overlap with other transactions, $ l_0 = r_0 = 0 $, $ L[0] = 0 $, and $ R[0] $ stores $ r(t_0, t_0) $ with $ O[0] = 0 $. For $ t_1 $, which overlaps with $ t_2 $ and $ t_3 $, $ l_1 = 1 $ and $ r_1 = 3 $, so $ L[1] = 1 $, and $ r(t_1, t_1) $, $ r(t_1, t_2) $, and $ r(t_1, t_3) $ are stored in $ R[1] $, $ R[2] $, and $ R[3] $, respectively, with $ O[1] = 1 $. For $ t_3 $, we query $ r(t_3, t_1) $ using the structure. It is stored at $ R[O[3] + 1 - L[3]] = R[6] $.

Compared with a matrix, the compact transitive closure reduces memory usage significantly. Its space complexity is linear to the number of results $r(t_i, t_j)$ that must be kept in the transitive closure rather than $O(|T|^2)$, where $|T|$ is the number of transactions. 

\noindent{\bf Computing Compact Transitive Closure.} The transitive closure of the known graph $G$ can be computed by Warshall's algorithm~\cite{CormenLR89} in $O(|V|^3)$ time. Notably, $G$ is a directed acyclic graph (DAG). The transitive closure of a DAG can be constructed by Purdom's algorithm~\cite{jr_transitive_1970} in $O(|E| + |V|^2)$ time. The algorithm consists of two steps: (1) obtaining a topological sort of vertices via depth-first search (DFS), and (2) maintaining the set of descendants $ D(v) $ for each vertex $ v $. Initially, $ D(v) = \{v\} $. Vertices are processed in reverse topological order, updating $ D(v) $ as $ D(v) \cup D(u) $ for each edge $ (v, u) \in E $.

In our problem, vertices represent transactions with timestamps. We improve Purdom's algorithm by leveraging temporal information to reduce its time complexity to $ O(|V| + |E \setminus \mathbb{T}|) $, where $ \mathbb{T} $ is the set of time-dependencies. The improved algorithm is detailed in Algorithm~\ref{alg:construct_trans_closure}.

\begin{algorithm}[t]
    \scriptsize
    \caption{Compact Transitive Closure Construction}
    \label{alg:construct_trans_closure}
    \begin{algorithmic}[1]
    \REQUIRE Graph $G = (V, E)$
    \ENSURE Compact transitive closure structure $(R, L, O)$
    \STATE $o_g \gets |V|$; $Q \gets$ an empty queue
    \FORALL{$t_i \in V$}
        \IF{$t_i$ is not visited}
            \STATE \texttt{DFS}$(t_i)$
        \ENDIF
    \ENDFOR
    \WHILE{$Q \neq \emptyset$}
        \STATE Dequeue $t_i$ from $Q$
        \STATE $s_i \gets \{t_i\}, d_i \gets r_i$ 
        \FORALL{$(t_i, t_j) \in E'$}
            \IF{$j \leq r_{m_i}$}
            \STATE $s_i \gets s_i \cup s_j $; $d_i \gets \min(d_i, d_j)$;
            \ENDIF
        \ENDFOR
        \STATE $j \gets r_i+1 $
        \WHILE{$j \leq r_{m_i}$}
                \STATE $s_i \gets s_i \cup s_j $; $d_i \gets \min(d_i, d_j)$;  $j \gets j+1 $
        \ENDWHILE
        \FOR{$t_j \in s_i$}
            \STATE $R[O[i] + j - L[i]] \gets \textbf{true}$
        \ENDFOR
        \FOR{$d_i \leq j \leq r_i$}
            \STATE $R[O[i] + j - L[i]] \gets \textbf{true}$
        \ENDFOR
    \ENDWHILE
    \RETURN $(R, L, O)$\\
    \textbf{Function} \texttt{DFS}($t_i$)
        \FOR{$(t_i, t_j) \in E'$}
            \IF{$t_j$ is not visited}
                \STATE \texttt{DFS}$(t_j)$
            \ENDIF
        \ENDFOR
        \FOR{$r_i < j < o_g$}
            \IF{$t_j$ is not visited}
                \STATE \texttt{DFS}$(t_j)$
            \ENDIF
        \ENDFOR
        \STATE Mark $t_i$ as visited
        \STATE Enqueue $t_i$ into $Q$
        \STATE $o_g \gets \min(o_g, r_i)$
    \end{algorithmic}
\end{algorithm}

To speed up the first step, we use temporal information to avoid unnecessary edge traversals during the DFS. Specifically, when a vertex $ t_i $ is visited, vertices $ t_j $ with $ j > r_i $ are already visited because $t_i$ reach these edges via time-dependency edges. A global variable $ o_g $ tracks the latest transaction whose descendants have been fully visited. When processing the successors of $t_i$, we only need to traverse the edges $(t_i, t_j)$ where $r_i < j < o_g$ or $(t_i, t_j) \in E \setminus \mathbb{T}$. After visiting $t_i$, $o_g$ is updated to $\min(o_g, r_i)$. Consequently, the time complexity of the first step of Purdom's algorithm is reduced to $O(|V| + |E \setminus \mathbb{T}|)$.

In the second step, we filter unnecessary edges and optimize the union of descendant sets. For three transactions $ t_i, t_j, t_k $, if $ (t_i, t_j) \in \mathbb{T} $ and $ (t_j, t_k) \in \mathbb{T} $, then $ (t_i, t_k) \in \mathbb{T} $. Therefore, it is unnecessary to update the descendant set $D(t_i)$ with $D(t_k)$ because $D(t_k)$ has already been merged into $D(t_j)$ according to the reverse topological sort, and $D(t_j)$ will be subsequently merged into $D(t_i)$. We define the minimum time-successor $ t_{m_i} $ of $ t_i $ such that $(t_i, t_{m_i}) \in \mathbb{T}$, and for all $(t_i, t_k) \in \mathbb{T}$, it holds that $\text{e}(t_{m_i}) \leq \text{e}(t_k)$. When updating $ D(t_i) $, edges $ (t_i, t_j) $ with $ j > r_{m_i} $ can be skipped.

To further optimize, we represent $ D(t_i) $ using two parts: an integer $ d_i $ such that all $ t_j $ with $ j > d_i $ are descendants, and a set $ S_i $ of remaining descendants with $ j \leq d_i $. Merging $ D(t_j) $ into $ D(t_i) $ updates $ d_i $ to $ \min(d_i, d_j) $ and $ S_i $ to $ S_i \cup S_j $. Since merging two sorted lists runs in $ O(d) $ time, the second step completes in $ O(d^2|V|) $.

\noindent{\bf Updating Compact Transitive Closure.} During constraint pruning and SAT problem solving, edges are added to the graph $ G = (V, E) $, potentially changing vertex reachability. Recomputing the transitive closure from scratch is inefficient for a small number of edge additions. Since $ G $ is a DAG, Italiano's algorithm~\cite{italiano_amortized_1986} can update the transitive closure in $ O(|V|) $ amortized time when an edge is added. We improve Italiano's algorithm by leveraging transactions' temporal information.

Italiano's algorithm identifies vertex pairs whose reachability changes due to the added edge $(t_i, t_j)$. If $ r(t_i, t_j) = \text{true} $, the edge has no effect. If $ r(t_j, t_i) = \text{true} $, it forms a cycle. Otherwise, the affected vertex pairs are:
\begin{equation}\label{eq:iset}
    \mathbb{I} = \{ (t_u, t_v) \mid r(t_u, t_i) = \text{true}, r(t_j, t_v) = \text{true}, r(t_u, t_v) = \text{false} \}.
\end{equation}
Italiano's algorithm visits all vertices to find $ \mathbb{I} $. We improve it by constructing a candidate set of vertex pairs based on transactions' temporal information, significantly reducing the search space. Specifically, we find:

\begin{lemma}\label{lem:u-candidate}
    For all $(t_u, t_v) \in \mathbb{I}$, we have $l_j \leq u \leq \min(r_i, r_j)$.
\end{lemma}
    


\begin{lemma}\label{lem:v-candidate}
    For all $(t_u, t_v) \in \mathbb{I}$, we have $\max(l_u, l_i, l_j) \leq v \leq \min(r_u, r_i)$.
\end{lemma}

Consequently, after adding an edge $(t_i, t_j)$ to $G$, we first obtain a set of candidate vertices for $t_u$ in $\mathbb{I}$ according to Lemma~\ref{lem:u-candidate}. For each candidate $t_u$, we obtain a set of candidate vertices for $t_v$ in $\mathbb{I}$ according to Lemma~\ref{lem:v-candidate}. If $t_u$ and $t_v$ satisfy the condition given by Eq.~\eqref{eq:iset}, $r(t_u, t_v)$ is updated to true.

\begin{figure}[t]
    \centering
    \subfigure[Graph $G$.]{
        \includegraphics[width=0.3\linewidth]{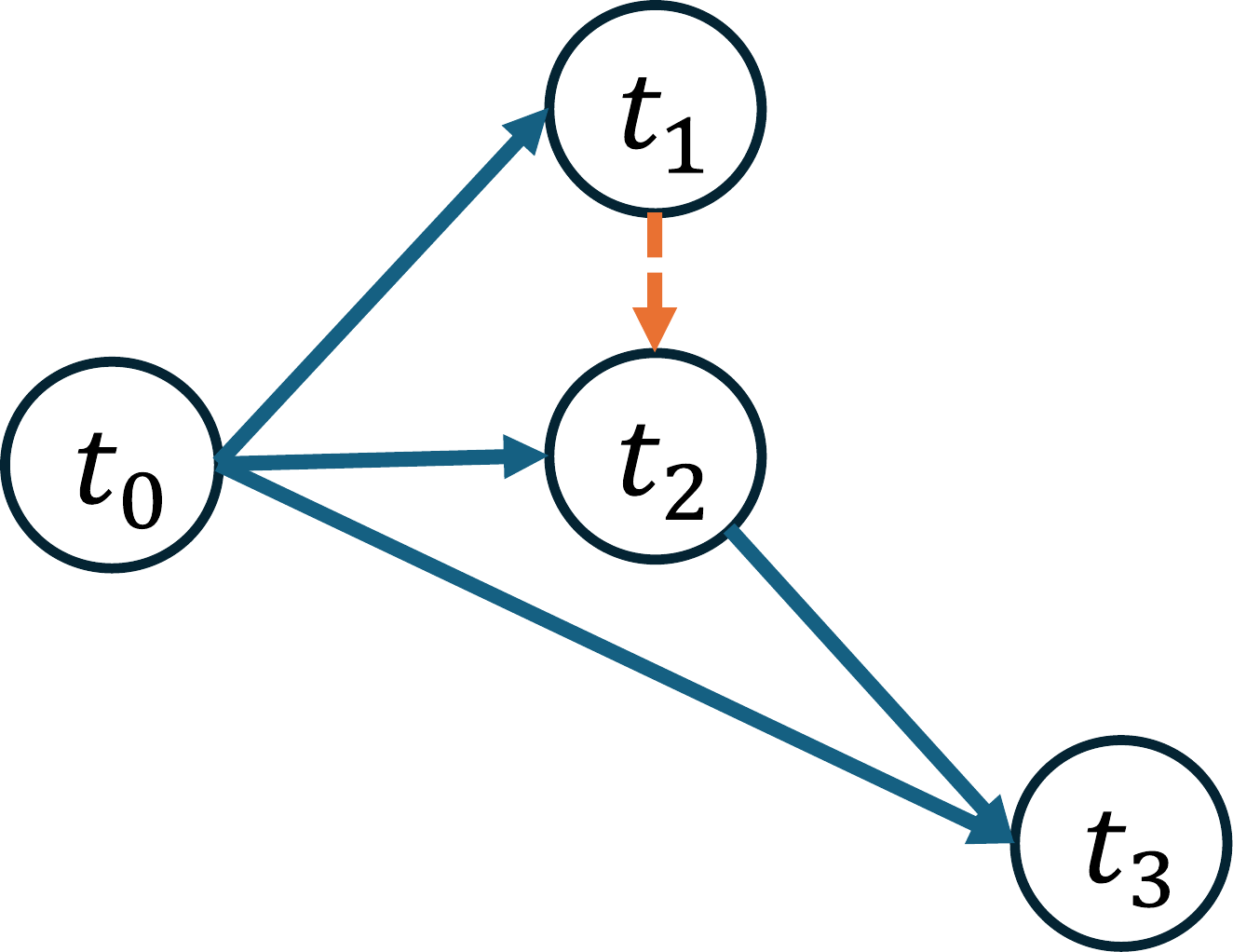}
        \label{fig:reach}
    }%
    \subfigure[Transitive closure $R$.]{
        \includegraphics[width=0.3\linewidth]{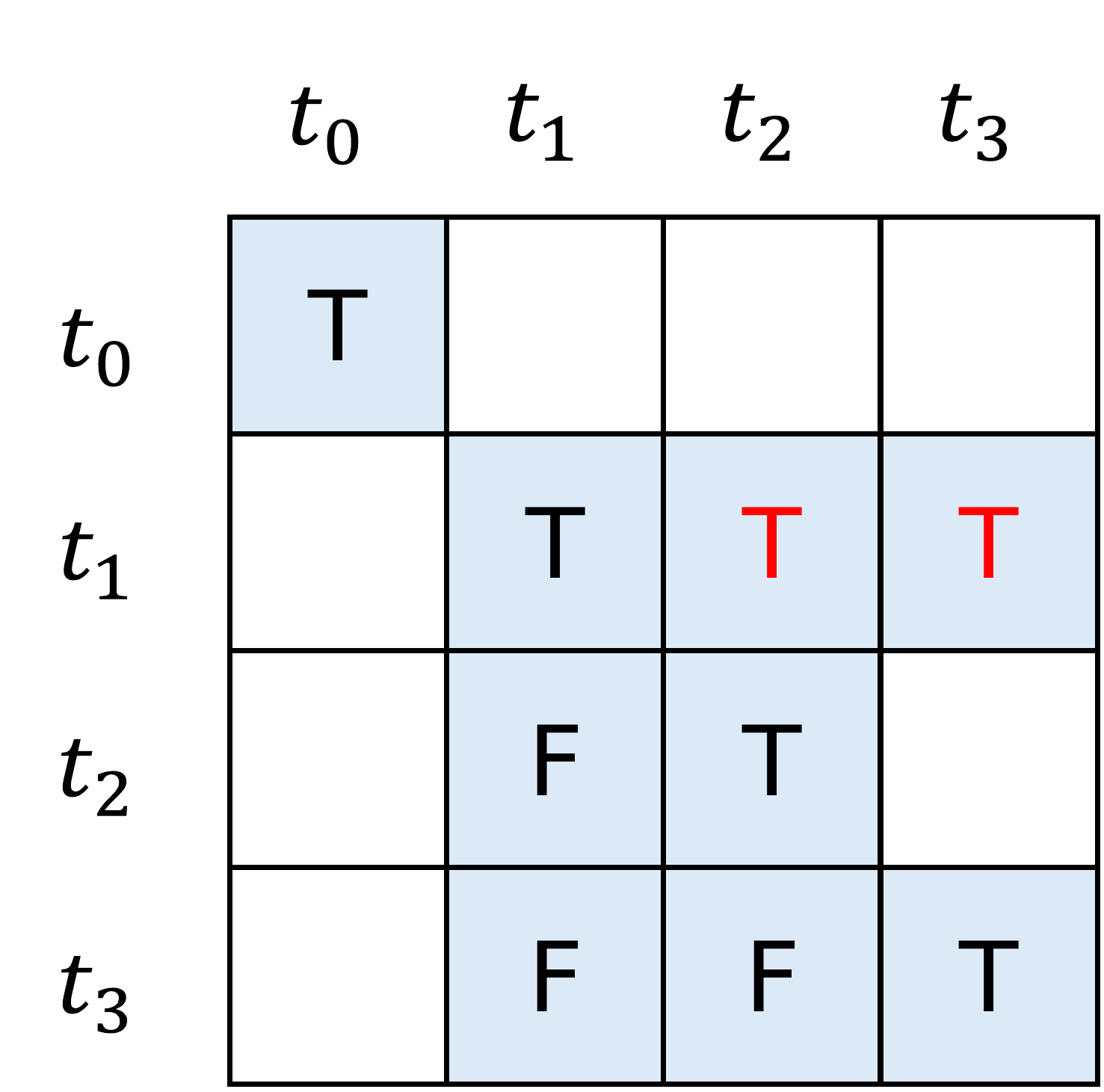}
        \label{fig:reachMatrix}
    }%
    \subfigure[Path matrix $P$.]{
       \includegraphics[width=0.3\linewidth]{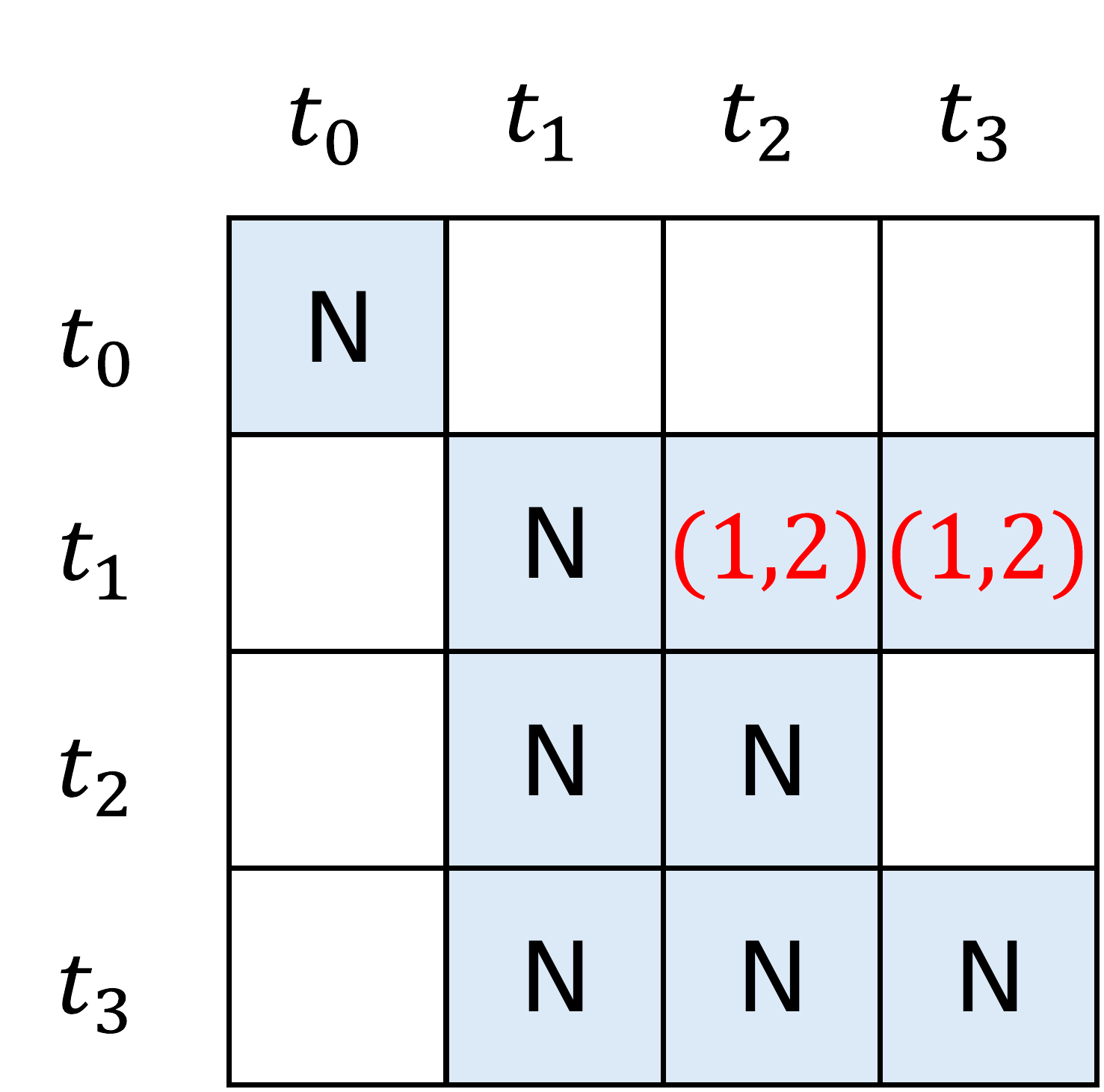}
       \label{fig:pathMatrix}
    }%
    \vspace{-1em}
    \caption{Updating transitive closure and path matrix.}
    \label{fig:updateMatrix}
    \vspace{-1em}
\end{figure}

Figure~\ref{fig:reach} shows the graph of the history in \autoref{fig:compact-transitive-closure}. When we add the edge $(t_1, t_2)$ to the graph, we start by finding the candidate set for $t_u$, which is $\{t_1, t_2, t_3\}$. For the candidate $t_1$, we compute the candidate set for $t_v$ and get $\{t_2, t_3\}$. We see that both $(t_1, t_2)$ and $(t_1, t_3)$ are in $\mathbb{I}$, so we update $r(t_1, t_2)$ and $r(t_1, t_3)$ to true. We repeat this process for the other candidates for $t_u$. Finally, we have the updated transitive closure shown in Figure~\ref{fig:reachMatrix}.

\noindent{\bf Path Finding.}
When solving the SAT problem, we need to find a path from transaction $t_i$ to $t_j$ if reachable. We maintain an array $P$ similar to $R$. When adding an edge $(t_p, t_q)$ makes $r(t_i, t_j)$ true, we update $P[O[i] + j - L[i]]$, i.e., $p(i,j)$, to $(t_p, t_q)$.

To retrieve the path from $t_i$ to $t_j$, we access $p(i,j)$ to obtain the edge $(t_p, t_q)$, which decomposes the path into $t_i \rightsquigarrow t_p \to t_q \rightsquigarrow t_j$. We recursively retrieve $p(i,p)$ and $p(q,j)$ to reconstruct the subpaths $t_i \rightsquigarrow t_p$ and $t_q \rightsquigarrow t_j$. This process continues until $p(m, n)$ directly corresponds to the edge $(t_m, t_n)$.

Figure~\ref{fig:pathMatrix} illustrates the update of $P$ when edge $(t_1, t_2)$ is added to the graph in Figure~\ref{fig:reach}. This addition sets $r(t_1, t_2)$ and $r(t_1, t_3)$ to T, updating $p(t_1, t_2)$ and $p(t_1, t_3)$ to $(t_1, t_2)$. To find the path from $t_1$ to $t_3$, we access $p(t_1, t_3)$, retrieve $(t_1, t_2)$, and decompose the path as $t_1 \to t_2 \rightsquigarrow t_3$. Since $p(t_2, t_3)$ is absent in $P$, it implies that $t_2$ reaches $t_3$ via the time-dependency edge $(t_2, t_3)$. Thus, we add $(t_2, t_3)$ to complete the path $t_1 \to t_2 \to t_3$.

\subsection{Simplified SAT Problem Formulation}
\label{sub:simplified-sat-problem-formulation}

\textsf{Cobra} introduces a boolean variable for each transaction pair operating on the same object and enforces mutual exclusivity through numerous constraints. We design a simplified formulation for item constraints and extend support to predicate constraints.

\noindent{\bf Variables.} Let $C^I$ be the set of item constraints. For each $c = (E_{ij}, E_{ji}) \in C^I$, we introduce a boolean variable $b_c$, where $b_c = T$ iff edges in $E_{ij}$ are added to $G$, and $b_c = F$ iff edges in $E_{ji}$ are added.

Let $C^P$ be the set of predicate constraints. Each $c \in C^P$ consists of edge sets ${ E_{\theta,t} \mid t \in F_{\theta,y} }$. We introduce a boolean variable $b_{c, t}$ for each $E_{\theta,t}$, where $b_{c, t} = T$ iff all determined edges in $E_{\theta,t}$ are added to $G$. Additionally, for each undetermined edge $e$, we introduce a boolean variable $b_e$, setting $b_e = T$ iff $e$ is added to $G$.

\noindent{\bf Literals.} Let $l_e$ denote the literal for an edge $e$, which is either a boolean variable or its negation, where $l_e = T$ iff $e$ is added to $G$: (1) For an undetermined edge $e$,  $l_e = b_e$. (2) For a determined edge $e$ in $E_{\theta,t}\in c$  set $l_e = b_{c, t}$. (3) For an edge $e$ in an item constraint $c=(E_{ij}, E_{ji})$, $l_e = b_c$ if $e \in E_{ij}$ and $l_e = \lnot b_c$ if $e \in E_{ji}$.

\noindent{\bf Formula.}
To ensure that exactly one edge set in a predicate constraint $c$ is added to $G$, we impose: 
\begin{equation}\label{eq:pred-known-edge}
    P_1 = \underbrace{\left(\bigvee_{E_{\theta,t} \in c} b_{c, t}\right)}_{\text{At least one set is chosen}} \land \underbrace{\left(\bigwedge_{E_{\theta,t}, E_{\theta,t'} \in c} (\neg b_{c, t} \lor \neg b_{c, t'})\right)}_{\text{No more than one set is chosen}}.
\end{equation}

An undetermined edge $e$ is added to $G$ only if the corresponding edge set $E_{\theta,t} \in c$ and the write-dependency edge $e'$ that derives $e$ are both added to $G$.This condition is formulated as:
\begin{equation}\label{eq:pred-unknown-edge}
    P_2 = b_e \lor \neg b_{c, t} \lor \neg l_{e'},
\end{equation}
where $b_{c, t}$ corresponds to $E_{\theta,t}$.

For serializability verification, we enforce the acyclicity of the generated compatible graph $G'$: 
\begin{equation}\label{eq:acyclic}
    P_3 = \text{acyclic}(G').
\end{equation}
Thus, serializability verification reduces to an SAT problem that checks whether an assignment satisfies $P_1 \land P_2 \land P_3$. If no such assignment exists, the history is not serializable; otherwise, it has an acyclic DSG.

\subsection{Customized SAT Problem Solver}
\label{sub:customized-sat-problem-solver}

\textsf{Cobra} uses MonoSAT~\cite{bayless_sat_2015} to solve the SAT problem. However, MonoSAT cannot handle our simplified SAT problem. Because MonoSAT requires a boolean variable to correspond to an edge, and in our formalization, a boolean variable may represent an edge set. Additionally, MonoSAT does not fully utilize the information existing in the transitive closure. We design a new solver for our simplified SAT problem formulation.

Our solver consists of a MiniSAT~\cite{EenS03} solver and a theory solver. MiniSAT is used to provide a partial assignment that satisfies the formula $P_1 \land P_2$. The theory solver checks whether $P_3$ is satisfied by the assignment provided by MiniSAT with help of the compact transitive closure. The theory solver also provides information to help MiniSAT generate assignments. There are two key components in this process, namely propagation and analysis.

\noindent{\bf Propagation.}
The propagation process assigns truth values for additional boolean variables based on the partial assignment provided by MiniSAT. It uses the pruning process (\S\ref{sub:constraint-reduction}) to choose edge sets or constraints need to be pruned and assign the corresponding variables to be $T$ or $F$.

\begin{figure}[t]
    \centering
    \vspace{-1em}
    \includegraphics[width=0.4\linewidth]{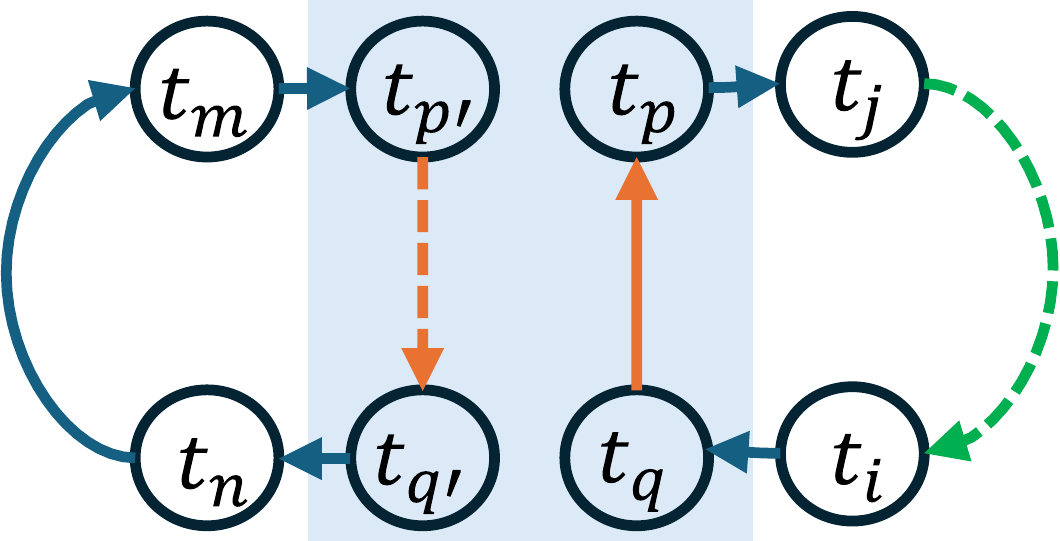}
    \caption{Solve Example}
    \label{fig:solver}
    \vspace{-1em}
\end{figure}

For example, in the known graph shown in \autoref{fig:solver}, where the blue solid arcs represent the edges already in the graph. Consider an item constraint $c = (E_{pq}, E_{qp})$, where $E_{pq} = \{ (t_{p'}, t_{q'})\}$ and $E_{qp} = \{ (t_{q}, t_{p})\}$. After adding edge $(t_m, t_n)$, $t_{p'}$ becomes reachable from $t_{q'}$, so adding $(t_{p'}, t_{q'})$ to the graph will form a cycle. Thus, we choose to add edge $(t_q, t_p) \in E_{qp}$ to the graph and assign $b_c = F$.

\noindent{\bf Analysis.}
Assignments provided during the propagation process may cause the known graph to contain cycles. The analysis process finds the reasons leading to the cycles and generates clauses to expand the formula $P_1\land P_2$ to prevent the same cycle for future assignments.

Let us proceed with the example in \autoref{fig:solver}. Suppose certain assignments set the literal for the edge $(t_j, t_i)$ to $T$. When we attempt to add $(t_j, t_i)$ to the known graph $G$, we find a path $p_{ij} = t_i \rightarrow t_q \rightarrow t_p \rightarrow t_j$ which forms a cycle with $(t_j, t_i)$. To avoid this cycle, we generate the following clause
\begin{equation*}
    p_1 = \neg l_{iq} \lor \neg l_{qp} \lor \neg l_{pj} \lor \neg l_{ji},
\end{equation*}
where $l_{iq}$, $l_{qp}$, $l_{pj}$, and $l_{ji}$ are the literals corresponding to edges $(t_i, t_q)$, $(t_q, t_p)$,  $(t_p, t_j)$, and $(t_j, t_i)$. This clause indicates that when all four literals are set to $T$, the four edges forming a cycle are added to $G$, rendering the clause $p_1$ unsatisfied. By adding this clause to $P_1 \land P_2$, future assignments that would cause the same cycle are avoided. In the implementation, we obtain deeper reasons by finding the First Unique Implication Point (F-UIP)~\cite{zhang_efficient_2001}. Deeper reasons can prevent more invalid assignments. 

\section{Evaluation}
\label{sec:evaluation}

We implemented our black-box serializability verification method \textsf{Vbox} in C++ and evaluated its performance by experiments. We compared Vbox with three existing verification methods \textsf{BE}~\cite{biswas_complexity_2019}, \textsf{Cobra}~\cite{tan_cobra_2020} and \textsf{Leopard}~\cite{li_leopard_2023}. All the experiments were conducted on a Linux server equipped with an Intel Xeon Gold 6130 CPU and 512GB of RAM. \textsf{Cobra} uses GPU to accelerate transitive closure computation, while other methods use CPU.

\subsection{Completeness}
\label{sub:completeness-evaluation}

We first evaluate the completeness of the verification methods in terms of their ability to detect various types of anomalies.

\noindent{\bf Workloads.}
To evaluate the completeness of the verifiers, we selected from~\cite{tan_cobra_2020} three real world transaction execution histories with 7 anomalies in serializability. We also synthesized workloads using the method in~\cite{li_coo_2022} and executed these workloads under the Read Uncommitted isolation level of MySQL to produce transaction execution histories containing 29 anomalies in serializability. These anomalies can be divided into three categories WAT, IAT, and RAT~\cite{li_coo_2022} which contain 13, 9 and 7 anomalies, respectively. 

\noindent{\bf Results.}
\autoref{tab:eval-completeness} shows the number of anomalies detected by the verification methods. It is remarkable that our method \textsf{Vbox} successfully detects all the real and the synthetic anomalies, while other methods fail to detect all the anomalies. This is because \textsf{Vbox} detects aborted reads and intermediate reads and can cover all the anomalies by item constraints and predicate constraints.

\begin{table}[t]
    \centering
    \scriptsize
    \caption{Fractions of detected anomalies.}
    \label{tab:eval-completeness}
    \vspace{-1em}
    \begin{tabular}{ccrrrr}
        \toprule
        History type & Anomaly type & \textsf{Vbox} & \textsf{Cobra} & \textsf{BE} & \textsf{Leopard} \\ 
        \midrule
        Synthetic & RAT & 13/13 & 6/13 & 7/13 & 3/13 \\ 
        Synthetic & WAT & 9/9 & 5/9 & 5/9 & 0/9 \\ 
        Synthetic & IAT & 7/7 & 5/7 & 5/7 & 0/7 \\ 
        Real-world & Real & 7/7 & 7/7 & 6/7 & 1/7 \\
        \bottomrule
    \end{tabular}
    \vspace{-1em}
\end{table}

\textsf{Cobra} and \textsf{BE} exhibit consistent anomaly detection capabilities since determining the commit order in \textsf{BE} is fundamentally equivalent to identifying the acyclic compatible graph in \textsf{Cobra}. Both \textsf{Cobra} and \textsf{BE} miss some anomalies because both of them do not check for aborted reads and intermediate reads, assume each transaction writes to each object only once, and lack support for predicates. These limitations are overcome by \textsf{Vbox}.

\textsf{Leopard} demonstrates the weakest anomaly detection capability among the three methods. This is mainly because that \textsf{Leopard} relies on the time information of the operations in the transactions, which is absent in the real-world histories. Although the synthetic histories contain time information, \textsf{Leopard} still fails to recognize some anomalies due to its incomplete abstraction of MySQL's concurrency control protocol, which only detects anomalies resulting from concurrent writes and does not address those caused by reads. \textsf{Vbox} also overcomes this limitation of \textsf{Leopard}.

\subsection{Efficiency}
\label{sub:efficiency-evaluation}

We also evaluate the efficiency of the verification methods in terms of their execution time and memory usage. 

\noindent{\bf Workloads.}
To evaluate the efficiency of different verification methods, we generated transaction execution histories using three benchmarks TPC-C~\cite{tpc_c}, C-Twitter~\cite{bigdata_twitter}, and BlindW~\cite{tan_cobra_2020}. These benchmarks were executed under the Serializable isolation level of PostgreSQL, generating 6 execution histories of 10,000 transactions named TPC-C, C-Twitter, BlindW-RH, BlindW-WR, Blind-WH, and BlindW-Pred, respectively.

(1) TPC-C is a standard OLTP benchmark that includes five types of transactions: new order, order status, payment, delivery, and stock level. We run the benchmark with one warehouse and keep the rest of the settings at their defaults.

(2) C-Twitter simulates transaction patterns in social networks and is designed for high-concurrency, frequently updated situations. We follow the implementation described in~\cite{tan_cobra_2020}.

(3) BlindW is a synthetic benchmark that features a randomly generated table with 10,000 rows and a schema of $(k, v_1, v_2)$. It performs both read-only and write-only transactions on this table. There are four variants of BlindW: BlindW-RH (80\% reads + 20\% writes), BlindW-WR (50\% reads + 50\% writes), BlindW-WH (20\% reads + 80\% writes), and BlindW-Pred (50\% predicate read and 50\% predicate write). The predicates are randomly generated range filters applied to either column $v_1$ or column $v_2$.

\begin{figure}[t]
    \centering
    \small
    \includegraphics[width=0.8\linewidth]{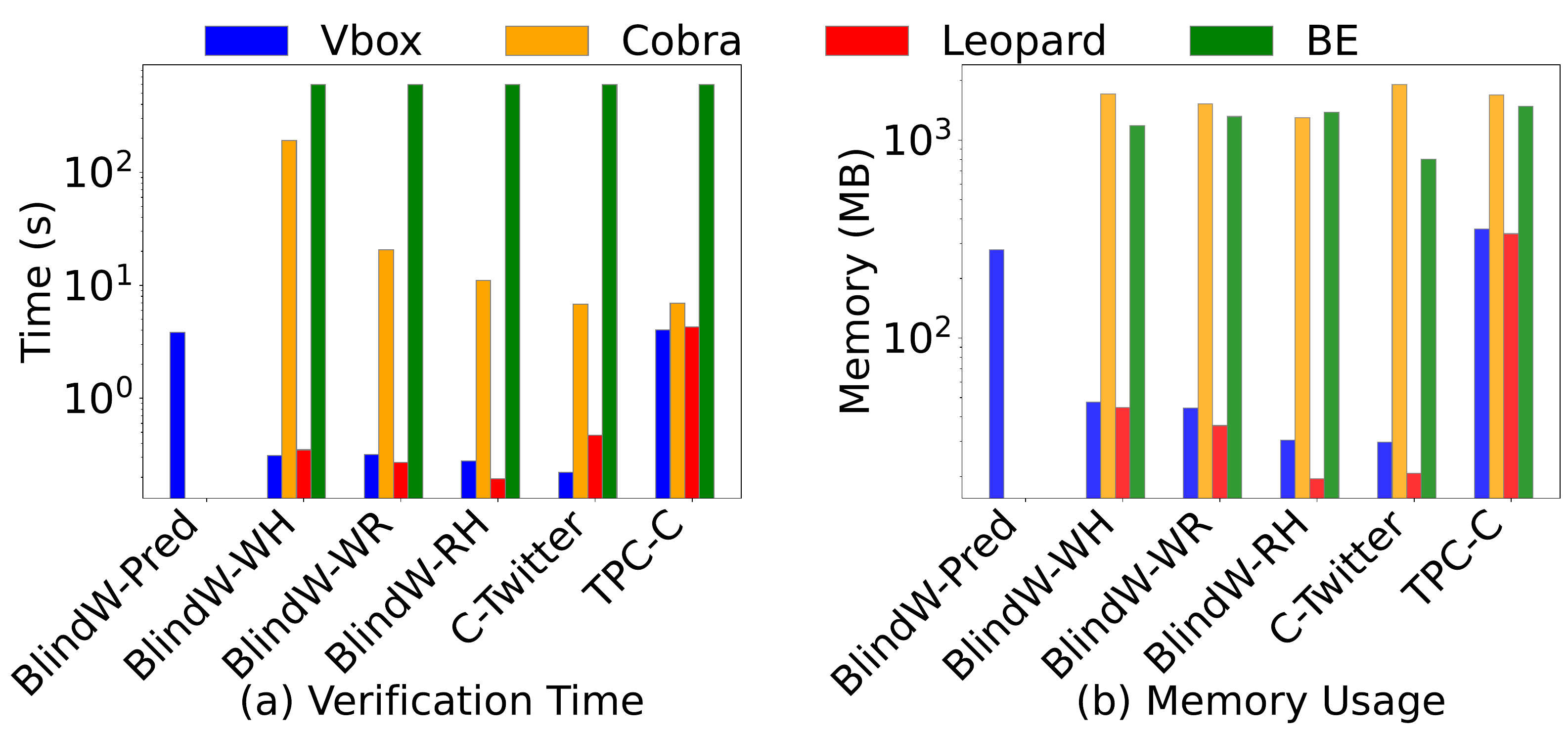}
    \caption{Verification time and memory usage.}
    \label{fig:efficiency-evaluation}
    \vspace{-2em}
\end{figure}

\noindent{\bf Results.}
\autoref{fig:efficiency-evaluation} shows the experimental results. On most workloads, our method \textsf{Vbox} has the shortest verification time because it employs a variety of techniques to enhance efficiency, including constraint pruning, constraint consolidation, improved graph algorithms based on time information, and the customized SAT solver. Additionally, the memory usage of \textsf{Vbox} is low as it maintains a compact transitive closure instead of the full transitive closure.

\textsf{BE} times out (\textgreater 10min) for all the workloads due to its exponential time complexity. \textsf{BE} exhaustively traverses all possible commit orders and stores a large number of intermediate states during the search, leading to significant memory consumption.

\textsf{Leopard} achieves low verification time and low memory usage comparable to those of \textsf{Vbox} because \textsf{Leopard} avoids constructing a DSG for the observed transaction history. Instead, it directly checks whether the history satisfies specific protocol requirements, resulting in linear time and space costs.

While both \textsf{Cobra} and \textsf{Vbox} verify serializability by constructing a compatible graph, Cobra uses Warshall's algorithm to compute and maintain the full transitive closure. Additionally, it applies an unsuitable SAT solver, which contributes to higher verification time and memory usage compared to \textsf{Vbox}.

\subsection{Scalability}
\label{sub:scalability-evaluation}

\begin{figure}[t]
    \centering
    \subfigure[Execution time.]{
        \includegraphics[width=0.4\linewidth]{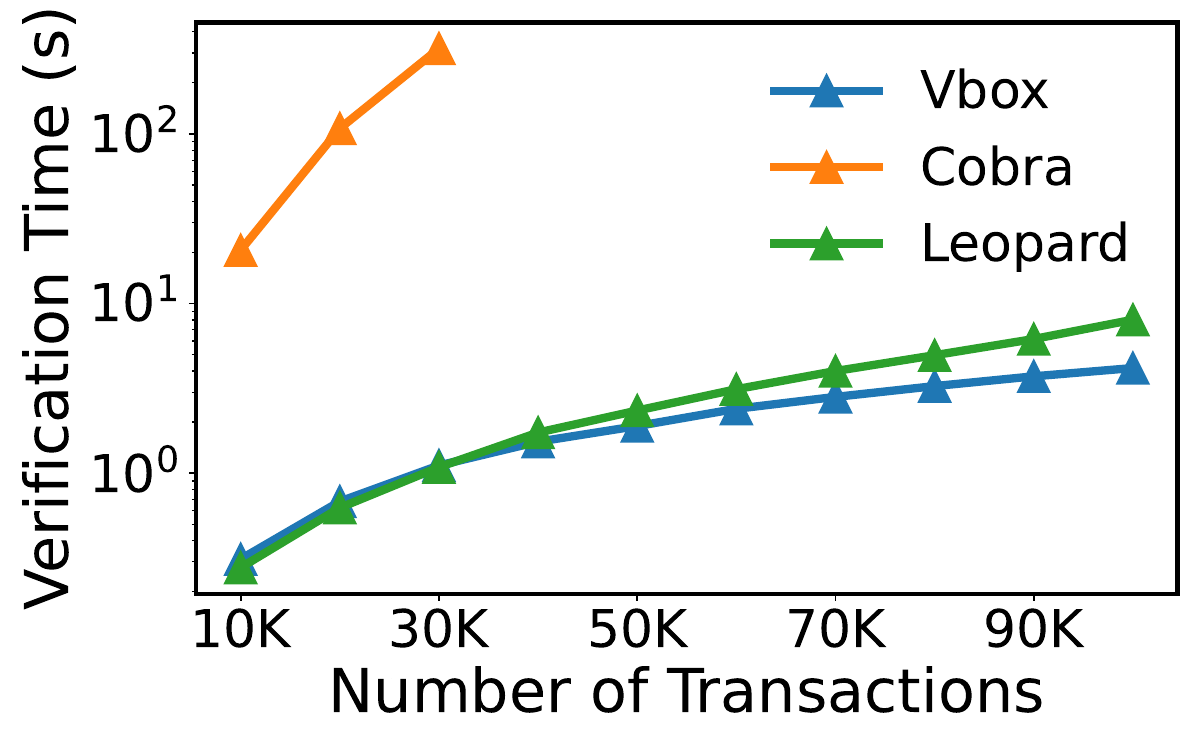}
        \label{fig:txn-num-time}
    }%
    \subfigure[Memory usage.]{
        \includegraphics[width=0.4\linewidth]{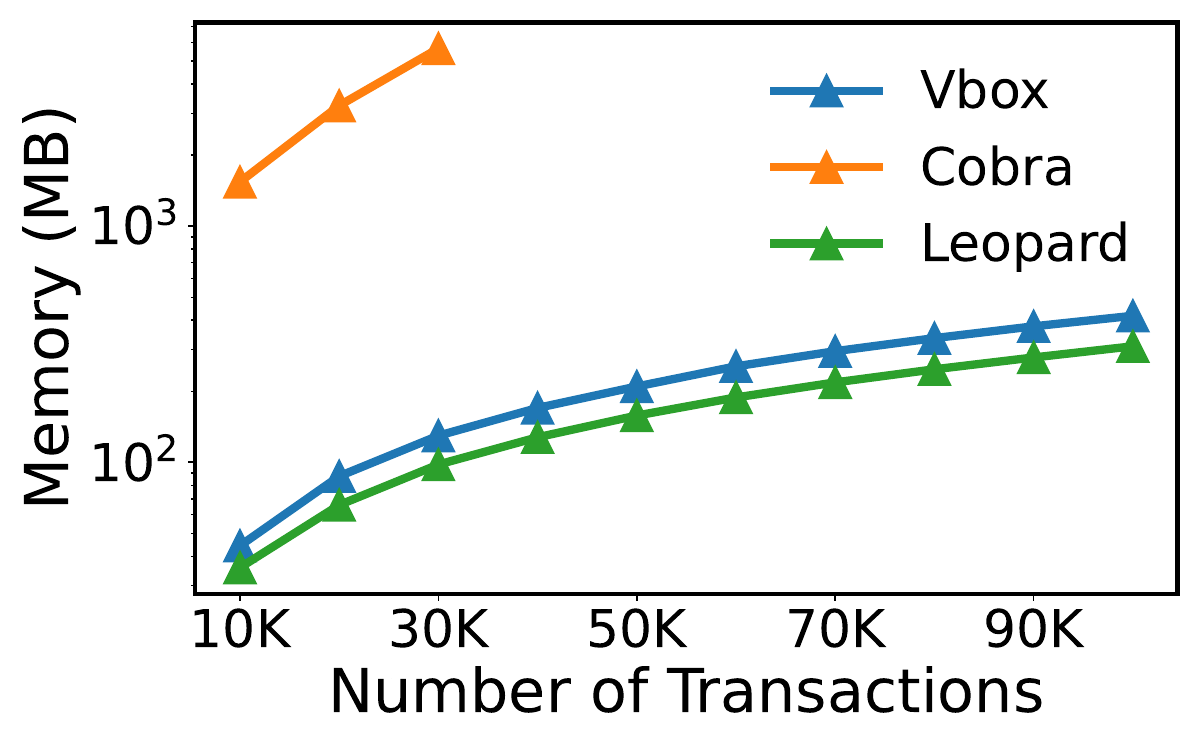}
        \label{fig:txn-num-overload}
    }%
    \vspace{-1em}
    \caption{Scalability.}
    \label{fig:scalability-evaluation}
    \vspace{-1em}
\end{figure}

Next, we evaluate the scalability of the verification methods with respect to the number of transactions in the history. We generated 10 BlindW-WR execution histories containing 10,000--100,000 transactions and verified these histories using different methods. The results are shown in \autoref{fig:scalability-evaluation}. Note that \textsf{BE} times out (\textgreater 10min) on the history composing of 10,000 transactions, so \textsf{BE} is not included in the results.

As the number of transactions in the history increases from 10K to 100K, the verification time of \textsf{Vbox} increases from 0.31s to 4.16s (13.4X), and the memory usage of \textsf{Vbox} increases from 44MB to 417MB (9.5X), exhibiting near-linear scalability. This is primarily attributed to usage of the compact transitive closure and the optimized algorithms for constructing and updating the closure.

\textsf{Leopard}'s verification time and memory usage also scale almost linearly. However, when the number of transactions exceeds 30,000, Leopard's verification time exceeds that of \textsf{Vbox}. This is because Leopard requires checking whether the operations overlap in time, and as the number of transactions grows, the cost of this checking process increases.

The scalability of \textsf{Cobra} is not good. As the number of transactions in the history increases from 10,000 to 30,000, the verification time of \textsf{Cobra} increases from 20s to 318s, and it times out given over 40,000 transactions. 

\subsection{Effectiveness}

Lastly, we evaluate the effectiveness of the techniques specially designed for \textsf{Vbox}.

\begin{figure}[t]
    \centering
    \subfigure[Construction.]{
        \includegraphics[width=0.4\linewidth]{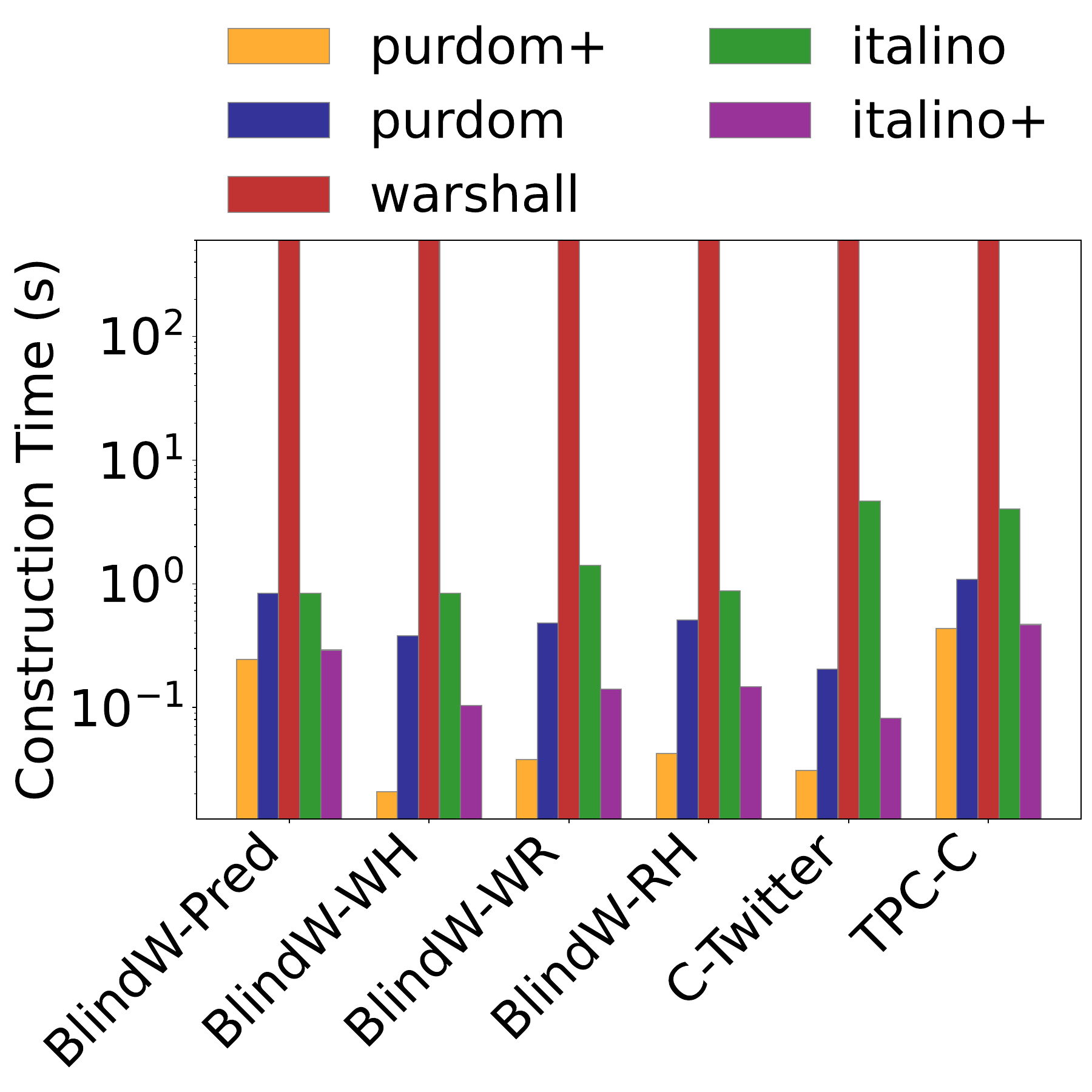}
        \label{fig:construct-closure}
    }%
    \subfigure[Updating.]{
    \includegraphics[width=0.4\linewidth]
    {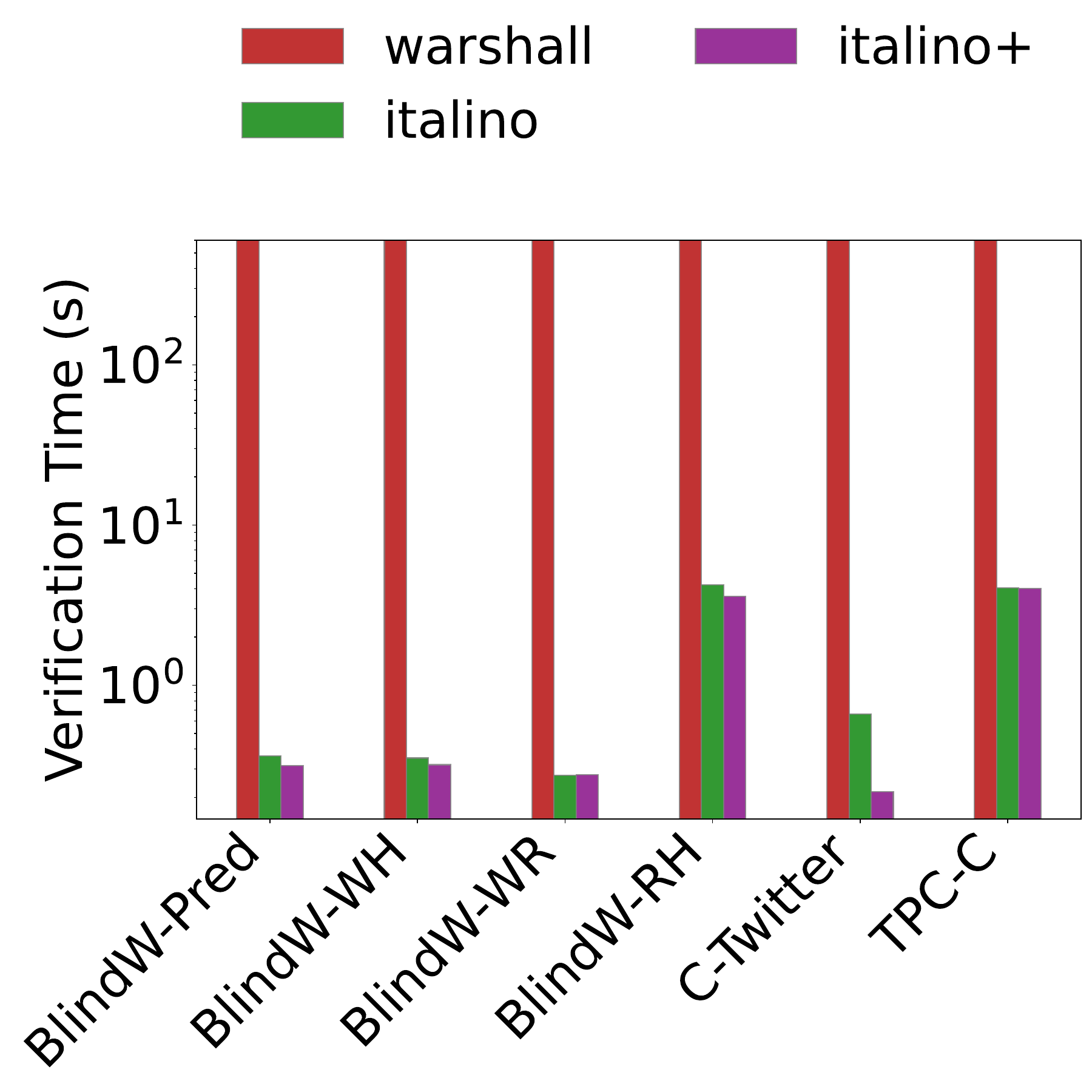}
        \label{fig:update-closure}
    }%
    \vspace{-1em}
    \caption{Constructing and updating transitive closure.}
    \label{fig:closure-evaluation}
    \vspace{-1em}
\end{figure}

\noindent{\bf Transitive Closure Construction.}
Figure~\ref{fig:construct-closure} shows the time for constructing the full transitive closure by Warshall's, Purdom's and Italino's algorithms and the time for constructing the compact transitive closure by our improved Purdom's (Purdom+) and our improved Italino's algorithm (Italino+). Note that Italino and Italino+ construct the transitive closure by adding edges one by one.

For all histories, Purdom+ achieves the fastest transitive closure construction, 3–-17X faster than the original Purdom's algorithm. This is because Purdom+ leverages transactions' time information to accelerate topological sorting and speed up merging descendant sets. Although Italiano's algorithm is primarily designed for incremental updating of the transitive closure, Italiano+ outperforms the original Purdom's algorithm. Warshall's algorithm times out (\textgreater 10min) for all histories because its time complexity is $O(n^3)$, where $n$ is the number of transactions.

\noindent{\bf Transitive Closure Updating.}
Figure~\ref{fig:update-closure} shows the verification time of different variants of \textsf{Vbox} using different transitive closure updating methods, including reconstructing the full transitive closure by Warshall's algorithm, updating the full transitive closure by Italino's algorithm and updating the compact transitive closure by Italino+. Obviously, \textsf{Vbox} with Italino+ achieves the shortest verification time because transactions' time information is used to filter out unnecessary updates to the transitive closure.

\noindent{\bf Constraint Reduction.}
\autoref{tab:eval-prune-num} compares the number of item constraints remaining after constraint reduction in \textsf{Vbox} and \textsf{Cobra}. The row named ``total'' records
the total number of potential item constraints between every pair of transactions that write the same object. Both \textsf{Vbox} and \textsf{Cobra} can reduce a great number of item constraints, and \textsf{Vbox} is more powerful than \textsf{Cobra} in constraint reduction because it adds time-dependency edges into the known graph, which enables constructing a more complete known graph and makes it possible to prune more useless constraints.

\begin{table}[t]
    \centering
    \scriptsize
    \caption{Quantity of item constraints after reduction.}
    \label{tab:eval-prune-num}
    \vspace{-1em}
    \begin{tabular}{lrrrrr}
        \toprule
        & BlindW-WH & BlindW-WR & BlindW-RH & C-Twitter & TPC-C \\ 
        \midrule
        Total & 211,566 & 98,451 & 25,908 & 530,605 & 612,565 \\ 
        \textsf{Cobra} & 17,061 & 3,829 & 437 & 1,371 & 0 \\ 
        \textsf{Vbox} & 43 & 17 & 2 & 40 & 0 \\ 
        \bottomrule
    \end{tabular}
    \vspace{-1em}
\end{table}

\noindent{\bf SAT Problem Solvers.}
Next, we compare our SAT solver VboxSAT with MiniSAT~\cite{EenS03} and MonoSAT~\cite{bayless_sat_2015}. VboxSAT can handle our simplified SAT problem formulation. For MonoSAT, a boolean variable is introduced for every pair of transactions operating on the same object, and the constraints are encoded by Eq.~\eqref{eq:cobra-sat-formulation}. For MiniSAT, two boolean variables $b_{ij}$ and $b_{ji}$ are introduced to represent dependencies $t_i \rightarrow t_j$ and $t_j \rightarrow t_i$, respectively. Acyclicity is enforced using the formula $(b_{ij} \land \neg b_{ji}) \lor (\neg b_{ij} \land b_{ji})$, which ensures that only one of $t_i \rightarrow t_j$ or $t_j \rightarrow t_i$ holds.

\begin{figure}[t]
    \centering
    \includegraphics[width=0.6\linewidth]{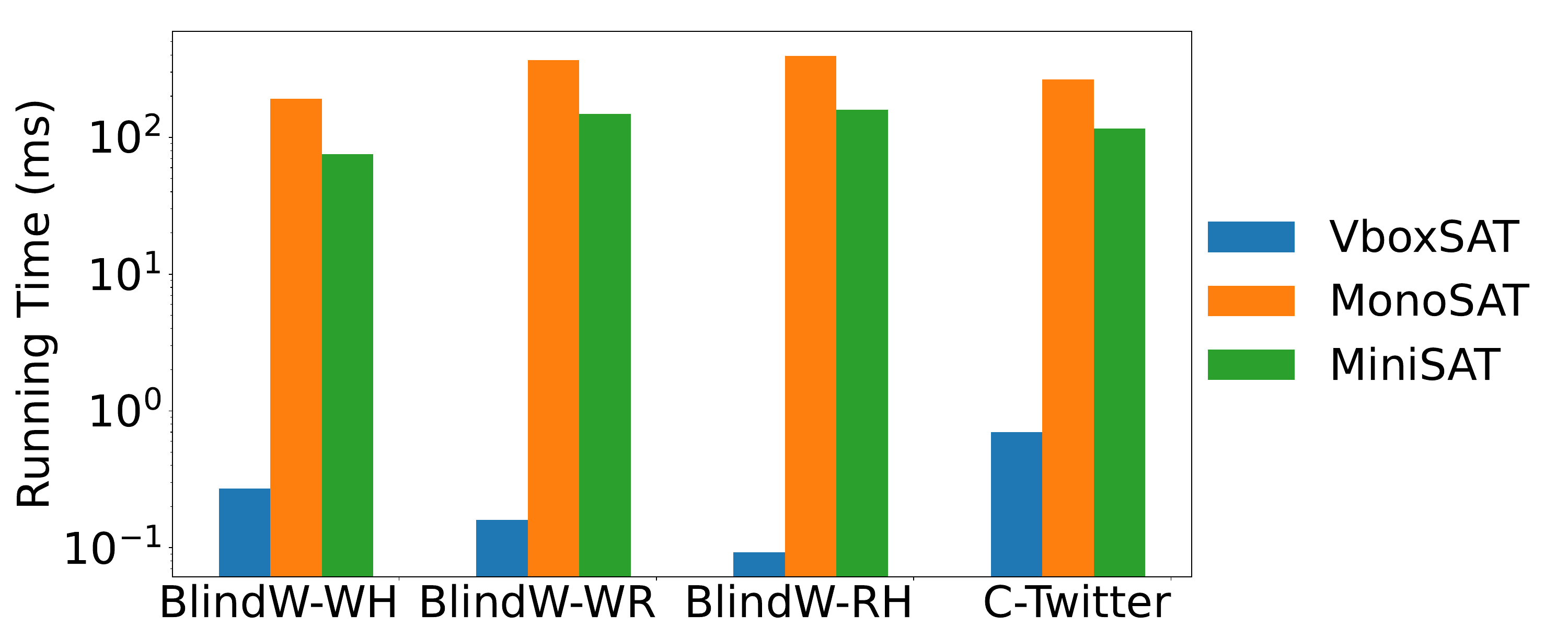}
    \caption{Running time of SAT solvers.}
    \label{fig:eval-solver}
    \vspace{-1em}
\end{figure}

\autoref{fig:eval-solver} shows the running time of the solvers, including the formula generation time. VboxSAT is 3--4 orders of magnitudes faster than MiniSAT and MonoSAT because VboxSAT significantly reduces the number of formulas and leverages known transitive closure information in detecting acyclicity without external formulas. MiniSAT is slow because it uses a large number of formulas to constrain acyclicity. MonoSAT is also slow because it must rebuild the graph and requires numerous formulas to describe the mutual exclusivity of the edge sets in item constraints.

\section{Conclusion}
\label{sec:conclusion}

\textsf{Vbox} is verfied to be correct and more efficient and more capable of detecting more data anomalies than the existing methods, while not relying on any specific concurrency control protocols. The superiority of \textsf{Vbox} in anomaly detection results from its capability of characterizing anomalies related to predicate read and write operations. The advantages of \textsf{Vbox} in time and space efficiency are attributed to the application of client-side time information of transactions in known graph construction and constraint reduction, the adoption of the compact transitive closure, and the efficient formulation and solver for the SAT problem. Future work includes developing an online version of \textsf{Vbox} for real-world production scenarios and extending \textsf{Vbox} to support the verification of weak isolation levels such as repeatable read.

\bibliographystyle{ACM-Reference-Format}
\bibliography{sample-base}


\end{document}